\documentclass[journal, ,x11names]{IEEEtran}

\usepackage{setspace}

\usepackage{subfigure}
\usepackage{amsmath,amsfonts}
\usepackage{algorithmic}
\usepackage{algorithm}
\usepackage{array}
\usepackage{subfig}
\usepackage{textcomp}
\usepackage{stfloats}
\usepackage{url}
\usepackage{verbatim}
\usepackage{graphicx}
\usepackage{cite}
\usepackage{subfloat}
\usepackage{tikz}
\usepackage{bm}

\usepackage{color,xcolor}
\usepackage{caption}

\usepackage{pgfplots}
\usetikzlibrary{calc,arrows.meta,positioning}
\usetikzlibrary{shapes.arrows}
\usetikzlibrary{fit,backgrounds} 
\usepackage{standalone}
\usepackage[numbers,sort&compress]{natbib}
\usepackage{setspace}
\usepackage{multirow}
\usepackage{makecell}

\begin{document}

\title{Context Video Semantic Transmission with Variable Length and Rate Coding over MIMO Channels}

\author{Bingyan Xie, Yongpeng Wu,~\IEEEmembership{Senior Member,~IEEE,} Wenjun Zhang,~\IEEEmembership{Fellow,~IEEE,} Derrick Wing Kwan Ng,~\IEEEmembership{Fellow, IEEE}, and Merouane Debbah,~\IEEEmembership{Fellow,~IEEE}

\thanks{(Corresponding author: Yongpeng Wu)}
\thanks{Bingyan Xie, Yongpeng Wu, and Wenjun Zhang are with the Department of Electronic Engineering, Shanghai Jiao Tong University, Shanghai 200240, China (e-mail:bingyanxie, yongpeng.wu, zhangwenjun@sjtu.edu.cn).}
\thanks{Derrick Wing Kwan Ng is with the School of Electrical Engineering and Telecommunications, University of New South Wales, Sydney, NSW 2052, Australia (e-mail: w.k.ng@unsw.edu.au).}
\thanks{M. Debbah is with KU 6G Research Center, Khalifa University of Science and Technology, P O Box 127788, Abu Dhabi, UAE (email: merouane.debbah@ku.ac.ae) and also with CentraleSupelec, University Paris-Saclay, 91192 Gif-sur-Yvette, France.}
}

\maketitle
\begin{abstract}
	The evolution of semantic communications has profoundly impacted wireless video transmission, whose applications dominate driver of modern bandwidth consumption. However, most existing schemes are predominantly optimized for simple additive white Gaussian noise or Rayleigh fading channels, neglecting the ubiquitous multiple-input multiple-output (MIMO) environments that critically hinder practical deployment. To bridge this gap, we propose the context video semantic transmission (CVST) framework under MIMO channels. Building upon an efficient contextual video transmission backbone, CVST effectively learns a context-channel correlation map to explicitly formulate the relationships between feature groups and MIMO subchannels. Leveraging these channel-aware features, we design a multi-reference entropy coding mechanism, enabling channel state-aware variable length coding. Furthermore, CVST incorporates a checkerboard-based feature modulation strategy to achieve multiple rate points within a single trained model, thereby enhancing deployment flexibility. These innovations constitute our multi-reference variable length and rate coding (MR-VLRC) scheme. By integrating contextual transmission with MR-VLRC, CVST demonstrates substantial performance gains over various standardized separated coding methods and recent wireless video semantic communication approaches. The code is available at \emph{https://github.com/xie233333/CVST}.
\end{abstract}


\section{Introduction}
The rapid proliferation of video-centric applications, e.g. virtual reality, Internet of Things, and live streaming, now constitutes major Internet traffic. These applications generate vast volumes of video data, imposing significant strain on existing wireless transmission systems. Conventionally, this challenge has been addressed through separated source-channel coding (SSCC) schemes, which integrate video codecs, e.g. H.265 and versatile video coding (VVC) \cite{265, vvc}, with channel codecs, e.g. low density parity check (LDPC), to compress and transmit video data. While SSCC schemes remain widely used due to their convenient modular design, the emerging deep learning (DL)-based joint source-channel coding (JSCC) schemes have demonstrated superior performances in finite blocklength regimes \cite{jscc, djscc}. Recent developments in JSCC have been particularly impactful in semantic communications \cite{mhsc,NTSCC,LCFSC,MU-LCFSC,genai,contrast,secure,dvc,wvsc,dvsc,dvst}, which have inspired video-specific frameworks. For example, Xie et al. \cite{wvsc} have introduced a semantic-level framework for efficiently exploring the frame relationship within group of pictures (GoPs). Niu et al. \cite{dvsc} have conducted signal-to-noise ratio (SNR)-adaptive channel coding with semantic restoration. In addition, Wang et al. \cite{dvst} have developed a context-based non-linear transform coding (NTC) framework to achieve video variable length coding (VLC). The context-based video transmission scheme \cite{dvst}, in accordance with the deep video coding \cite{dcvc, dv-c, fvc}, offers greater compressibility than residual-based schemes \cite{wvsc, dvsc} and enable multi-reference awareness schemes. During video transmission, \cite{dvst} generates the context, which is a compact representation of spatial-temporal information, as side information for fusing with current frame features as codewords for entropy coding and transmission.

\begin{table*}[htbp]
	\centering
	\caption{Comparison of Existing Video Semantic Communication Frameworks and the Proposed CVST Scheme.}
	\label{table1}
	
	\begin{tabular}{|c|c|c|c|c|c|c|}  
		\hline
		& & & & & &\\[-6pt] 
		Framework&Type&VLC&VRC&Channel-aware backbone&Rate-aware backbone&Design for MIMO\\
		\hline
		& & & & & &\\[-6pt] 
		CVST (Ours)&Context-based&$\surd$&$\surd$&$\surd$&$\surd$&$\surd$\\
		\hline
		& & & & & &\\[-6pt]  
		WVSC \cite{wvsc}&Residual-based&$\times$&$\times$&$\surd$&$\times$&$\times$\\
		\hline
		& & & & & &\\[-6pt]  
		DVSC \cite{dvsc}&Residual-based&$\times$&$\times$&$\surd$&$\times$&$\times$\\
		\hline
		& & & & & &\\[-6pt] 
		DVST \cite{dvst}&Context-based&$\surd$&$\times$&$\times$&$\times$&$\times$\\
		\hline
	\end{tabular}
\end{table*}

While existing wireless video transmission schemes have demonstrated promising performances, several critical challenges persist. Some approaches \cite{dvsc, wvsc} adopt fixed-rate transmission, inherently lacking support for VLC. \cite{dvst} incorporates VLC with rate adaptation, it does not fully exploit channel-related information in context generation, entropy coding, and rate control, which leads to suboptimal rate–distortion performance in wireless video transmission. Notably, supplemental enhancement information (SEI) is critical for unlocking additional coding gains in conventional video coding \cite{film,face}. Motivated by this principle, in JSCC-based video transmission, instantaneous channel-related information (e.g., CSI and SNR) should be treated as SEI and embedded throughout the transmission pipeline to enable channel-aware semantic coding. Moreover, most existing video semantic transmission schemes \cite{dvsc,wvsc,dvst} evaluate robustness under simplified single-antenna channel models (e.g., AWGN and Rayleigh fading), thereby overlooking the complexity and prevalence of practical MIMO deployments. Recent MIMO semantic transmission works \cite{flsc, semimo} employ singular value decomposition (SVD) precoding together with DL-based channel estimation, feature arrangement, or joint detection \cite{score}, demonstrating the effectiveness of SVD for MIMO semantic transmission. Since SVD converts a MIMO channel into multiple parallel subchannels with unequal singular values (and thus different effective SNRs), the idea of context–channel pairing naturally arises; consequently, MIMO CSI-aware network design should be explicitly incorporated into video semantic transmission.

Moreover, existing frameworks \cite{dvc, dvsc, wvsc, dvst} typically require training multiple models for different rate points, which incurs substantial training and storage overhead. This limitation highlights the need for a once-trained variable rate coding (VRC) model. Recent efforts in variable-rate image semantic transmission offer promising directions \cite{swinjscc, padc, ntscc+}. For example, Ke et al. \cite{swinjscc} have proposed a spatial modulation module that scales latent representations based on the targeted rate. Zhang et al. \cite{padc} have established a VRC-enabled DeepJSCC along with a channel bandwidth ratio (CBR) optimizer given a PSNR quality constraint. Though achieving VRC results, such schemes \cite{swinjscc,padc} constrain transmission rates by regulating specific feature dimension for the encoder output layer, which is less flexible than the NTC-based scheme \cite{NTSCC}. As a result, Wang et al. \cite{ntscc+} have focused on integration between the NTC and feature modulation terms \cite{dcvc-hem}, which helps flexibly adjust the transmission rate. Similar to the quantization parameter, the feature modulation offers more flexible CBR control than feature dimension selection schemes. However, extending these techniques to wireless video transmission requires further innovation to develop CBR-aware network architectures.

Based on the above analysis, the absence of SEI-aware video transmission frameworks supporting variable length and rate coding (VLRC) motivates our proposal of context video semantic transmission (CVST), a novel once-trained video semantic communication framework enabling simultaneous channel and rate adaptation. The distinctions of CVST against other frameworks are outlined in Tab. \ref{table1}. To be specific, context is first extracted based on spatial-temporal information of the current frame. Then, a context-channel correlation map is learned to represent the MIMO channel effect brought to wireless video transmission. Leveraging the learned map, extracted context and motion vector are compressed by the proposed multi-reference entropy coding. To enable rate adaptivity, checkerboard feature modulation terms are subsequently integrated as quantization parameters for VRC. All the channel-state and rate-related contents are finally embedded as SEI into a unified rate allocation coder for joint adaptation. The contributions of this paper are summarized as follows  
\begin{enumerate}
\item{}
To effectively match MIMO channel characteristics and video features, we introduce a learnable context–channel correlation map. Inspired by cosine-similarity-based alignment \cite{clip}, the proposed map aligns group-divided context features with MIMO subchannels, enabling CSI-aware semantic feature allocation and robust adaptation to channel variations.
\item{}
To enhance both performance and practicality, we propose a multi-reference entropy-coding design for VLRC video semantic transmission. For VLC, we employ checkerboard-based entropy coding that leverages spatial, temporal, and channel-state references to enable efficient and parallel-friendly rate allocation across feature groups, while accounting for the distinct roles of motion and context. For VRC, we further introduce feature-modulation terms as rate-control parameters, enabling multiple rate points within a single trained model.
\item{}
To achieve channel-and rate-aware coding, we propose a multi-reference fusion coder for joint motion vector and context processing. It integrates three critical components: SNR values, the context-channel correlation map, and checkerboard modulation terms, enabling flexible adaptation to instantaneous channel conditions along with targeted CBRs. This synergistic integration ensures that both channel states and bandwidth constraints are embedded throughout the rate adaptation pipeline, facilitating SEI-aware rate allocation for time-varying wireless environments.
\item{}
To validate the effectiveness of CVST, we conduct comprehensive benchmarking against state-of-the-art separated coding schemes (VVC/H.265 + 5G NR LDPC), and leading DL-based frameworks \cite{dvsc, dvst}. Quantitative results demonstrate significant performance gains of CVST in both fixed-rate and variable-rate coding scenarios under MIMO channels.

\end{enumerate}

The rest of this paper is organized as follows. Section II introduces the framework overview of CVST. Section III illustrates the framework details. Section IV describes the deployment of CVST. Section V demonstrates the superiority of the proposed networks through a series of experiments. Section VI concludes the paper.

Notations: $\mathbb{R}$ and $\mathbb{C}$ refer to the real and complex number sets, respectively. $\mathcal{CN}\left (\mu, \sigma^2 \right)$ denotes a complex Gaussian distribution with mean $\mu$ and variance $\sigma^2$. $\lfloor\cdot \rceil$ denotes the quantization operation. $\oslash$ denotes the element-wise division operation while $\otimes$ denotes the element-wise multiplication operation. $\left(\cdot\right)^{H}$ denotes the Hermitian, $\left(\cdot\right)^{-1}$ is the matrix inverse, $\log(\cdot)$ denotes the logarithm operation.

\begin{figure*}[htbp]
	\centering
	\includegraphics[width=6.8in]{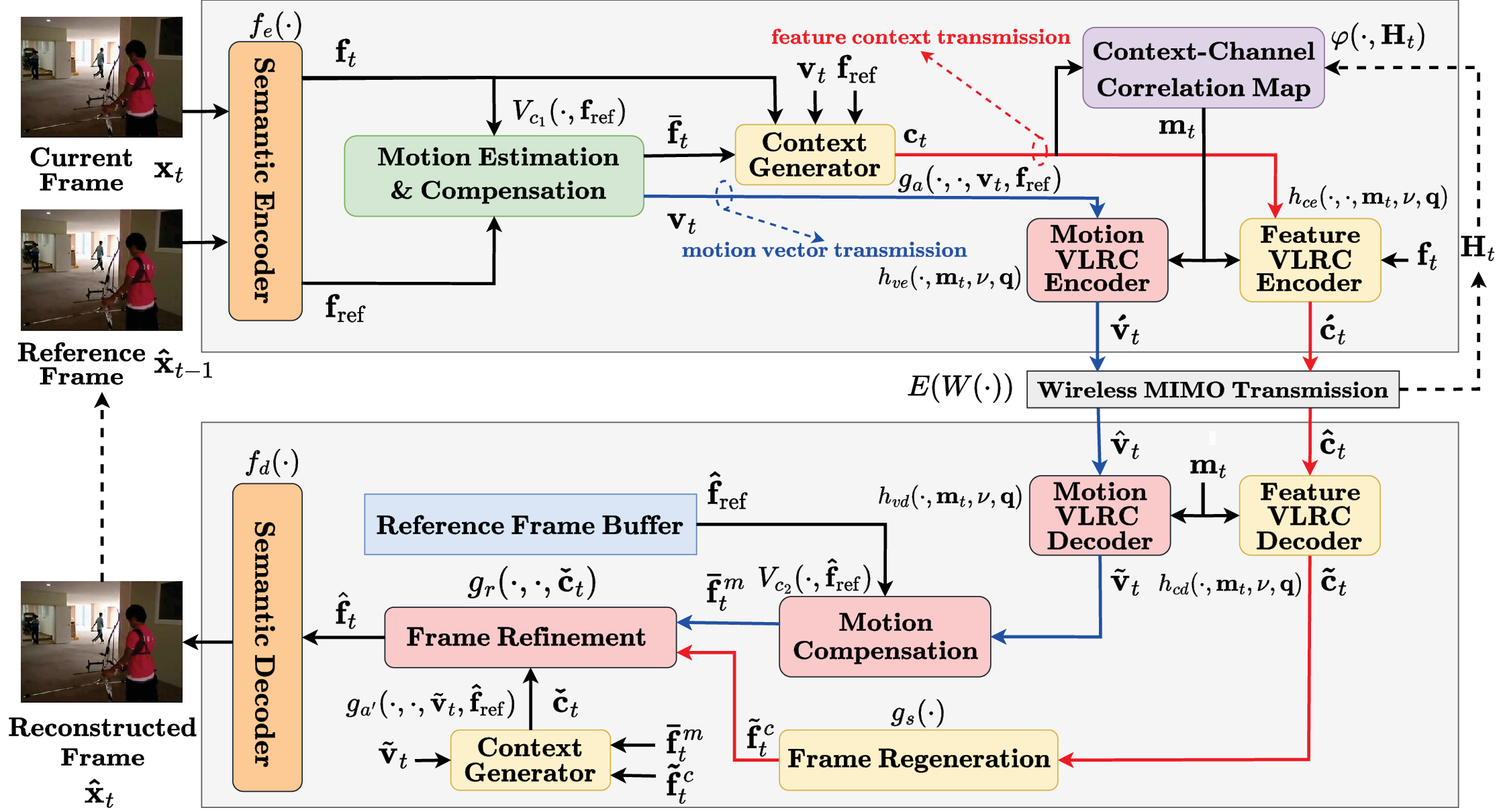}
	\caption{The proposed CVST framework. The red line is the context-based semantic feature transmission link while the blue line is the motion vector transmission link. The dashed line refers to the MIMO CSI feedback link.}
	\label{fig_1}
\end{figure*}

\section{Overview of proposed framework}
In this section, we consider a typical GoP-based wireless video semantic communication problem, similar to \cite{dvst}, and describe an overview of the proposed CVST framework.

The proposed CVST framework is shown in Fig. \ref{fig_1}. For arbitrary GoP $\mathbf{X} = \left\{\mathbf{x}_1,\mathbf{x}_2,\cdots,\mathbf{x}_T\right\}$, it contains $T$ successive frames with $\mathbf{x}_i\in \mathbb{R}^{3\times H\times W}$, $i=1,2,\cdots, T$. $\mathbf{x}_1$ is the intra-coded frame (I frame) which is transmitted to the receiver using image semantic communication schemes. The semantic encoder, $f_e(\cdot): \mathbb{R}^{3\times H\times W}\longrightarrow \mathbb{R}^{L\times H'\times W'}$, encodes $\mathbf{x}_t$ into the semantic features $\mathbf{f}_t$, with respective semantic frame dimension $L$, $H'$, and $W'$. Then, the motion vectors $\mathbf{v}_t\in \mathbb{R}^{L\times H'\times W'}$ and the predicted semantic frame $\mathbf{\bar{f}}_t\in \mathbb{R}^{L\times H'\times W'}$ are learned through the motion estimation $\&$ compensation module $V_{c1}(\cdot,\mathbf{f}_{\mathrm{ref}}):\mathbb{R}^{L\times H'\times W'}\longrightarrow \mathbb{R}^{L\times H'\times W'}$, where $\mathbf{f}_{\mathrm{ref}}=\mathbf{f}_{t-1}$. After that, with the context generator $g_a(\cdot,\cdot,\mathbf{v}_t,\mathbf{f}_{\mathrm{ref}}):\mathbb{R}^{L\times H'\times W'}\times\mathbb{R}^{L\times H'\times W'}\longrightarrow \mathbb{R}^{L\times H'\times W'}$, the context $\mathbf{c}_t \in \mathbb{R}^{L\times H'\times W'}$ is learned through both $\mathbf{\bar{f}}_t$ and $\mathbf{f}_t$. 

We aim to explore the correlation between wireless channel states and transmitted codewords. With both learned context and feedback MIMO CSI, we generate a context-channel correlation map $\mathbf{m}_t \in \mathbb{R}^{(L/N_t/m)\times N_t\times N_\mathrm{t}}$ through $\varphi(\cdot,\mathbf{H}_t):\mathbb{R}^{L\times H'\times W'}\longrightarrow\mathbb{R}^{(L/N_t/m)\times N_\mathrm{t}\times N_t}$ to represent its correlation, where $m$ refers to the feature channel numbers for a single group in a channel-wise division manner, $N_t$ refers to the MIMO transmitting antenna number. Then $\mathbf{m}_t$ is fed into the motion VLRC encoder $h_{ve}(\cdot, \mathbf{m}_t,\nu,\mathbf{q}): \mathbb{R}^{L\times H'\times W'}\longrightarrow \mathbb{R}^{L_{v}}$ and context VLRC encoder $h_{ce}(\cdot,\cdot, \mathbf{m}_t,\nu,\mathbf{q}): \mathbb{R}^{L\times H'\times W'}\times \mathbb{R}^{L\times H'\times W'}\longrightarrow \mathbb{R}^{L_{c}}$ to achieve the CSI-aware semantic coding along with variable length and rate adaptation for generating robust and flexible semantic codewords, $\mathbf{\acute{v}}_t \in \mathbb{R}^{L_{v}}$ and $\mathbf{\acute{c}}_t \in \mathbb{R}^{L_{c}}$, in terms of various CBRs. $L_{v}$ and $L_{c}$ are the respective final transmitted video codeword lengths. $\nu$ refers to the channel SNR value, while $\mathbf{q}$ refers to a series of feature modulation terms implying the transmission rates.

$\mathbf{\acute{v}}_t$ and $\mathbf{\acute{c}}_t$ are then reshaped and precoded by the SVD as
\begin{align}
	\{\mathbf{\hat{v}}_t,\mathbf{\hat{c}}_t\}=\Lambda^{-1}\mathbf{U}^H\mathbf{H}\mathbf{V}\{\mathbf{\acute{v}}_t,\mathbf{\acute{c}}_t\}+\Lambda^{-1}\mathbf{U}^H\mathbf{n}, 
\end{align}
where $\mathbf{H}\in \mathbb{C}^{N_r\times N_t}$ denotes the MIMO channel matrix, $N_r$ refers to the MIMO receiving antenna number, $\mathbf{n}$ is the complex Gaussian channel noise vector whose component has zero mean and covariance $\sigma^{2}$. SVD decomposes the acquired MIMO channel matrix $\mathbf{\tilde{H}}$. $\mathbf{\tilde{H}}=\mathbf{U}\Lambda\mathbf{V}^H$ with $\mathbf{U}\in \mathbb{C}^{N_r\times N_r}$, $\mathbf{V}\in \mathbb{C}^{N_t\times N_t}$ and $\Lambda\in \mathbb{R}^{N_r\times N_t}$.

At the receiver, with the motion VLRC decoder $h_{vd}(\cdot, \mathbf{m}_t,\nu,\mathbf{q}): \mathbb{R}^{L_{v}}\longrightarrow \mathbb{R}^{L\times H'\times W'}$ and context VLRC decoder $h_{cd}(\cdot, \mathbf{m}_t,\nu,\mathbf{q}): \mathbb{R}^{L_{c}}\longrightarrow \mathbb{R}^{L\times H'\times W'}$, received motion vector and context are translated to $\mathbf{\tilde{v}}_t$ and $\mathbf{\tilde{c}}_t$ with the help of context-channel correlation map. Through the motion compensation $V_{c2}(\cdot,\mathbf{f}_\mathbf{ref}):\mathbb{R}^{L\times H'\times W'}\longrightarrow \mathbb{R}^{L\times H'\times W'}$, frame regeneration $g_s(\cdot):\mathbb{R}^{L\times H'\times W'}\longrightarrow \mathbb{R}^{L\times H'\times W'}$ and context generator $g_a'(\cdot,\cdot,\mathbf{\hat{v}}_t,\mathbf{\hat{f}}_{\mathrm{ref}}):\mathbb{R}^{L\times H'\times W'}\times\mathbb{R}^{L\times H'\times W'}\longrightarrow \mathbb{R}^{L\times H'\times W'}$, $\bar{\mathbf{f}}^m_{t}$, $\tilde{\mathbf{f}}^c_t$ and $\check{\mathbf{c}}_t$ are learned to reconstruct the semantic frame $\hat{\mathbf{f}_t}$ with frame refinement $g_r(\cdot,\cdot,\check{\mathbf{c}}_t):\mathbb{R}^{L\times H'\times W'}\times\mathbb{R}^{L\times H'\times W'}\longrightarrow \mathbb{R}^{L\times H'\times W'}$. Finally, the semantic decoder, $f_d(\cdot):\mathbb{R}^{L\times H'\times W'}\longrightarrow \mathbb{R}^{3\times H\times W}$, converts $\hat{\mathbf{f}}_t$ into the final reconstructed GoP $\hat{\mathbf{X}}=\left\{\hat{\mathbf{x}}_1,\hat{\mathbf{x}}_2,\cdots,\hat{\mathbf{x}}_T\right\}$ frame by frame.

In this paper, we consider the typical low delay pattern (LDP) video transmission, which contains one I frame with many inter-coded frames (P frames) in a GoP. Since the I frame transmission is well studied, we specifically focus on the P frame wireless transmission over MIMO fading channels. The end-to-end CVST procedure is summarized in Alg. 1.

\begin{algorithm}[htbp]
	\caption{Wireless Video Transmission for CVST}\label{alg:alg2}
	\begin{algorithmic}
		\STATE 
		$\textbf{Input:}$ original $\left\{\mathbf{x}_1,\cdots,\mathbf{x}_T\right\}$ for a GoP
		\STATE
		$\textbf{Output:}$ reconstructed $\left\{\hat{\mathbf{x}}_1,\cdots,\hat{\mathbf{x}}_T\right\}$ for a GoP
		
		
		\STATE \hspace{0.5cm}$ \textbf{} $
		\STATE1. Acquire $\hat{\mathbf{x}}_1$ through wireless image transmission.
		\STATE2. \textbf{for $t=2,\cdots,T$ do}
		\STATE3. \hspace{0.2cm} Semantic encoding: $\mathbf{x}_t \xrightarrow{f_e(\cdot)} \mathbf{f}_t$.
		\STATE4. \hspace{0.2cm} Motion estimation $\&$ compensation, context generation:
		\STATE \hspace{0.55cm} $\mathbf{f}_t \xrightarrow{V_{c1}(\cdot,\mathbf{f}_{\mathrm{ref}}),g_a(\cdot,\cdot,\mathbf{v}_t,\mathbf{f}_{\mathrm{ref}})} \{\mathbf{v}_t,\mathbf{c}_t\}.$
		\STATE5. \hspace{0.2cm} VLRC encoding for motion vectors and context: 
		\STATE \hspace{0.55cm} $\{\mathbf{v}_t,\mathbf{c}_t\}\xrightarrow{h_{ve}(\cdot, \mathbf{m}_t,\nu,\mathbf{q}),h_{ce}(\cdot,\cdot, \mathbf{m}_t,\nu,\mathbf{q})}\{\mathbf{\acute{v}}_t,\mathbf{\acute{c}}_t\}$.
		\STATE6. \hspace{0.2cm} Wireless MIMO channel transmission as Eq. (1).
		\STATE7. \hspace{0.2cm} VLRC decoding for motion vectors and context: 
		\STATE \hspace{0.55cm} $\{\mathbf{\hat{v}}_t,\mathbf{\hat{c}}_t\}\xrightarrow{h_{vd}(\cdot, \mathbf{m}_t,\nu,\mathbf{q}),h_{cd}(\cdot, \mathbf{m}_t,\nu,\mathbf{q})}\{\mathbf{\tilde{v}}_t,\mathbf{\tilde{c}}_t\}$.
		
		\STATE8. \hspace{0.2cm} Motion compensation, frame regeneration:	
		\STATE \hspace{0.55cm} $\{\mathbf{\tilde{v}}_t,\mathbf{\tilde{c}}_t\}\xrightarrow{V_{c2}(\cdot, \mathbf{f}_\mathrm{ref}),g_s(\cdot)}\{\bar{\mathbf{f}}^m_t,\tilde{\mathbf{f}}^c_t\}$.
		\STATE9. \hspace{0.2cm} Context generation, frame refinement:
		\STATE \hspace{0.55cm} $\{\bar{\mathbf{f}}^m_t,\tilde{\mathbf{f}}^c_t\}\xrightarrow{g_a'(\cdot, \cdot,\mathbf{\hat{v}}_t,\mathbf{\hat{f}}_{\mathrm{ref}}),g_r(\cdot,\cdot,\check{\mathbf{c}}_t)}\hat{\mathbf{f}}_t$.
		\STATE10. \hspace{0.05cm} Semantic decoding: $\hat{\mathbf{f}}_t \xrightarrow{f_d(\cdot)} \mathbf{\hat{x}}_t$.
		\STATE11. \textbf{end for}
		
	\end{algorithmic}
	\label{alg1}
\end{algorithm}

\section{Details of CVST Framework}
In this section, we present the structure and detailed designs of each module in the CVST framework.

\subsection{Context-Channel Correlation Map}
\begin{figure}[htbp]
	\centering
	\includegraphics[width=3.4in]{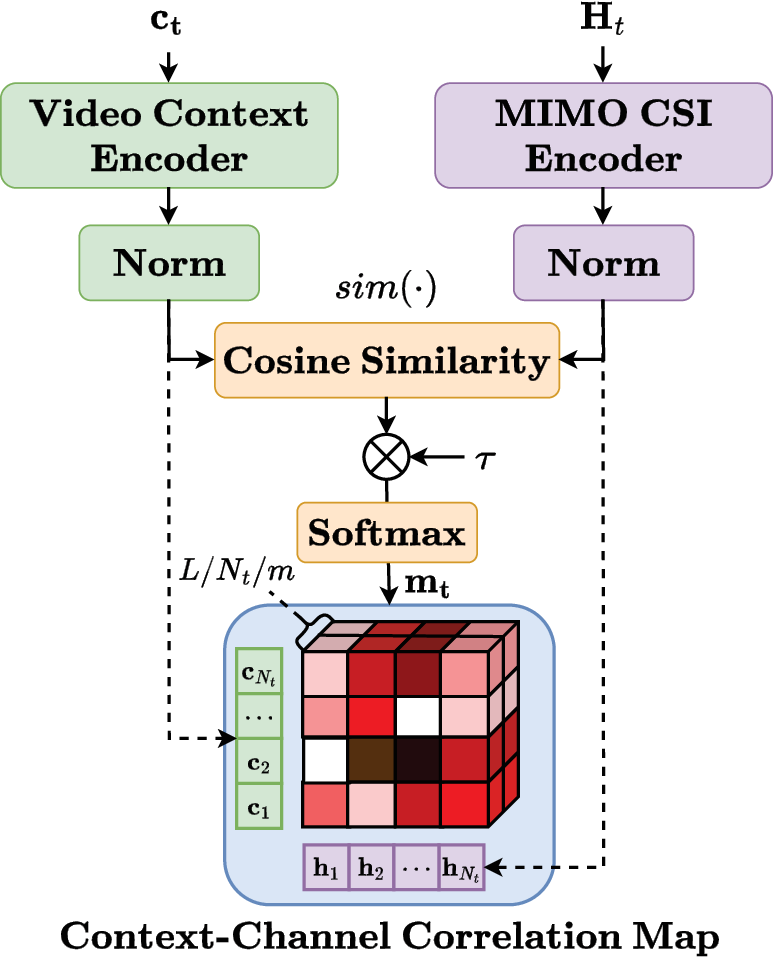}
	\caption{The context-channel correlation map for matching the context channel group and MIMO subchannel pairs.}
	\label{fig_2}
\end{figure}

As mentioned in Sec. \uppercase\expandafter{\romannumeral2}, we construct a context-channel correlation map to model interdependence between video context information and given MIMO CSI. Since extracted context provides compact spatiotemporal representations of video frames, such context-channel pair matching map effectively characterizes MIMO fading channel impacts on wireless video transmission. Crucially, different context channels exhibit varying semantic significance for reconstruction quality due to semantic differences among divided spatial regions. We therefore partition context into feature groups along the channel dimension. Correspondingly, SVD precoding decomposes MIMO channels into $N_t$ parallel subchannels with distinct noise intensity levels dictated by singular values. Inspired by CLIP's cross-modal alignment approach \cite{clip}, we compute pair-wise cosine similarity to establish correlations between context feature groups and MIMO subchannels. As shown in Fig. \ref{fig_2}, the video context $\mathbf{c}_t$ and feedback MIMO CSI $\mathbf{H}_t$ are mapped to the identical feature space through respective feature encoders and normalization. Then, the cosine similarity is computed to characterize the correlation among different context-channel pairs in the form of the context-channel correlation map $\mathbf{m}_t \in \mathbb{R}^{(L/N_t/m)\times N_t\times N_t}$.
It is note that the video feature context is divided along with the channel dimension $L$ whose each group has $m$ adjacent channels while MIMO channels are divided into $N_t$ subchannels with the $N_t\times N_r$ channel type. Take a specific case as an example, for $m=4$, $L=64$, and $8\times8$ MIMO channels, each $m_{t,ij}$ represents the matched score between the $i$-th channel group ($i=1,\cdots,N_t\times2$) with adjacent 4 feature channels and the $j$-th MIMO subchannel ($j=1,\cdots,N_t$) after the SVD precoding. Note that for $m=4$, there are 16 channel groups aligning with 8 subchannels. In this way, total 16 feature channel groups are divided into 2 gathers for each gather obtaining 8 feature channel groups to match the correlation with 8 MIMO subchannels, which can be concatenated later. The process for formulating the context-channel correlation map is shown as \cite{clip}
\begin{align}
m_{t,ij}(\mathbf{c}_{t},\mathbf{H}_{t}) = \frac{\exp(sim(\mathbf{V}_{\theta_1}(\mathbf{c}_{t,i}),\mathbf{V}_{\theta_2}(\mathbf{H}_{t,j}))/\tau)}{\sum_{j=1}^{N_t}\exp(sim(\mathbf{V}_{\theta_1}(\mathbf{c}_{t,i}),\mathbf{V}_{\theta_2}(\mathbf{H}_{t,j}))/\tau)},
\end{align}
where $m_{t,ij}(\mathbf{c}_{t},\mathbf{H}_{t})$ represents the score for providing the relative ranking of matched context-channel pair with the $i$-th context channel group, the $j$-th MIMO subchannel, $\mathbf{V}_{\theta_1}(\cdot)$ and $\mathbf{V}_{\theta_2}(\cdot)$ encapsulate the corresponding feature encoder and normalization process for video context and MIMO CSI, respectively. The video context encoder adapts convolutional neural network (CNN)-based structure \cite{cnn} while MIMO CSI encoder adapts the attention-based structure \cite{transformer}. $sim(\cdot)$ represents the cosine similarity computation, $\tau$ is the learnable temperature parameter for adjusting the scaling extent with a default value 0.07. After the group division, $\mathbf{m}_t$ can be reshaped into $\mathbf{m}_t\in\mathbb{R}^{(L/m)\times N_t}$.

There are several advantages for constructing such context-channel correlation map. First, it establishes a configurable bridge between feature contexts and MIMO subchannels, explicitly modeling source-channel interactions during JSCC transmission. Second, it enables the proposed CVST to intrinsically fuse the MIMO CSI into variable length/rate coding through side information embeddings, enhancing robustness against channel variations. Third, its CLIP-inspired softmax normalization constrains feature group scores to unity per subchannel, enabling the exact rate allocation in the channel group level for subsequent entropy coding in dynamic wireless environments.

\subsection{Multi-reference Entropy Coding for Variable Length and Rate Video Transmission}
To provide flexibility and practicality for the proposed CVST, we then illustrate the multi-reference entropy coding for variable length and rate video transmission with the help of previously learned context-channel correlation map step by step.

\begin{figure*}[htbp]
	\centering
	\includegraphics[width=7.0in]{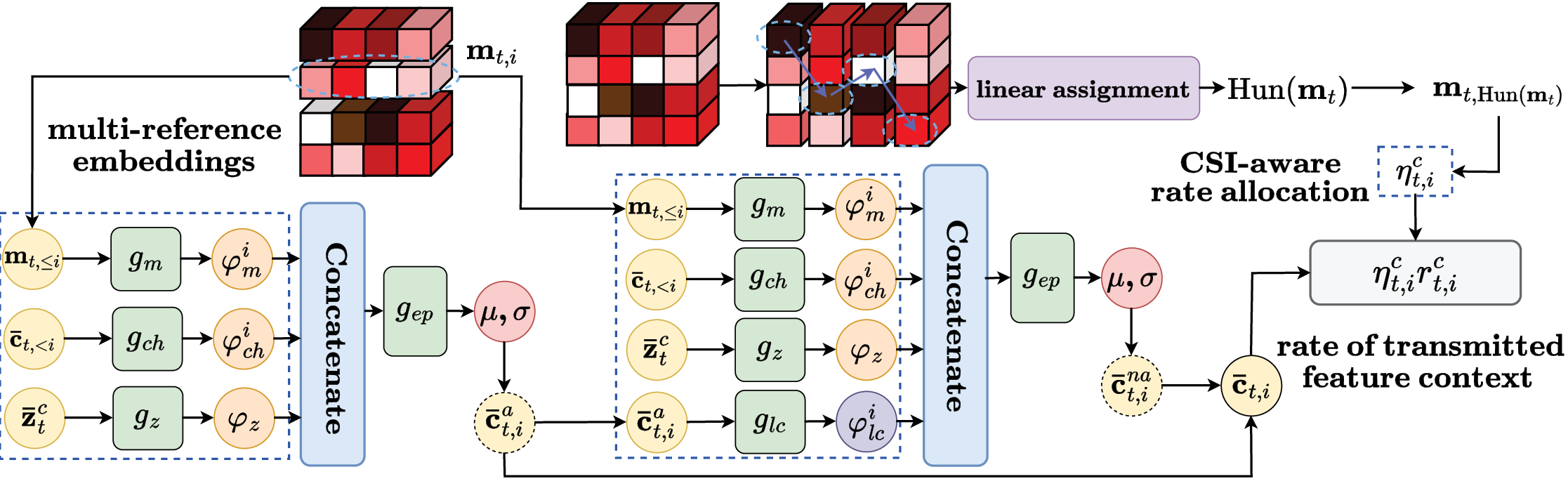}
	\caption{The proposed multi-reference variable rate and length entropy coding. The entropy coding is divided into anchored and non-anchored parts. For a specific feature channel group, the anchored part is first learned with multi-references embeddings. Then, the non-anchored part is learned. $\eta_{t,i}^c$ is adjustable according to the CSI-aware rate allocation.}
	\label{fig_3}
\end{figure*}
\subsubsection{Variable Length Coding for Wireless Video Transmission}
We first exploit the variable length coding in CVST. Similar to \cite{dvst}, we employ the NTC \cite{ntc} to achieve the variable length coding for finding the suitable rate point in the rate distortion curve rather than set the encoder output channel dimension fixed to achieve a predefined CBR. Since existing NTC schemes in \cite{dvst} follow the auto-regressive manner to conduct the entropy coding, the computation redundancy can be huge. The checkerboard model \cite{checkerboard}, which is widely adopted in the DL-based image compression, provides parallel-friendly performances for conducting entropy coding algorithms. Furthermore, inspired by \cite{mlic, dcvc-dc}, multiple references work as SEI to pose positive effect to ensure the accurate estimation of entropy distribution. We thus split the input semantic features along the feature channel dimension according to the learned context-channel correlation map for a group-wise checkerboard entropy coding. It ensures multiple references for conducting entropy coding of each slice since different semantic features have different spatial and temporal characteristics. Take the feature context as an example, the entropy of each feature context group for the anchored part is given as
\begin{align}
	r_{t,i}^{c,a} & = -\log P_{\bar{\mathbf{c}}_{t,i}^{a}|\mathbf{m}_{t,\le i},\bar{\mathbf{z}}_t^c,\bar{\mathbf{c}}_{t,<i}}(\bar{\mathbf{c}}_{t,i}^{a}|\mathbf{m}_{t,\le i},\bar{\mathbf{z}}_t^c,\bar{\mathbf{c}}_{t,<i}),
\end{align}
where $\bar{\mathbf{z}}_t^c= \lfloor \mathbf{z}_t^c\rceil=\lfloor h_r^c(\mathbf{c}_t)\rceil$ denotes the quantized hyperprior parameters for entropy coding, $\bar{\mathbf{c}}_{t,i}^{a}$ denotes the quantized anchored feature context of the $i$-th group. $\bar{\mathbf{c}}_{t,\le i}$ denotes the previously quantized feature context. Since we consider the group-based optimization, multiple references are embedded as conditions for conducting the context-based entropy coding. $\bar{\mathbf{z}}_t^c$ is considered as the hyperprior feature reference. $\mathbf{m}_{t,\le i}$ and $\bar{\mathbf{c}}_{t,<i}$ are the channel state reference and previous feature reference, respectively. $r_{t,i}^{c,a}$ is learned by the hyperprior entropy model containing both hierarchical prior and spatial prior.

Different from Eq. (5), for the non-anchored context, previously computed anchored feature context is employed as reference information containing spatial locality. It thus can be similarly formulated as
\begin{align}
	r_{t,i}^{c,na} & = -\log P_{\bar{\mathbf{c}}_{t,i}^{na}|\mathbf{m}_{t,\le i},\bar{\mathbf{z}}_t^c,\bar{\mathbf{c}}_{t,<i},\bar{\mathbf{c}}_{t,i}^a}(\bar{\mathbf{c}}_{t,i}^{na}|\mathbf{m}_{t,\le i},\bar{\mathbf{z}}_t^c,\bar{\mathbf{c}}_{t,<i},\bar{\mathbf{c}}_{t,i}^a).
\end{align}

The anchored and non-anchored parts of each slice remain adjacent in spatial domain. We then collect them together and present the complete entropy as 
\begin{align}
	r_{t}^c & = \sum_{i=1}^{m}(r_{t,i}^{c,a}+r_{t,i}^{c,na}) = \sum_{i=1}^{m}r_{t,i}^{c}.
\end{align}

In order to introduce the entropy model into the whole CVST training procedure, a uniform noise is injected into the feature context to replace the quantized representation $\bar{\mathbf{c}}_{t,i}$ as $\tilde{\mathbf{c}}_{t,i}=\mathbf{c}_{t,i}+\mathbf{o}_{t,i}$ with $\mathbf{o}_{t,i}\sim \mathcal{U}(-\frac{1}{2},\frac{1}{2})$ to enable backpropogation during training. Each $\tilde{\mathbf{c}}_{t,i}$ is modeled as a Laplace distribution with learned mean value $\tilde{\mu}_{t,i}$ and variance value $\tilde{\sigma}_{t,i}$. The polished entropy model is formulated as
\begin{equation}
	\begin{aligned}
		&P_{\tilde{\mathbf{c}}_{t}^{a}|\mathbf{m}_{t},\tilde{\mathbf{z}}_t^c}(\tilde{\mathbf{c}}_{t}^{a}|\mathbf{m}_{t},\tilde{\mathbf{z}}_t^c)\\&=\prod_{i}P_{\tilde{\mathbf{c}}_{t,i}^{a}|\mathbf{m}_{t,i},\tilde{\mathbf{z}}_t^c,\tilde{\mathbf{c}}_{t,<i}}(\tilde{\mathbf{c}}_{t,i}^{a}|\mathbf{m}_{t,i},\tilde{\mathbf{z}}_t^c,\tilde{\mathbf{c}}_{t,<i})\\&=\prod_{i}(\mathcal{L}(\tilde{\mu}_{t,i}^{c,a},\tilde{\sigma}_{t,i}^{c,a})*\mathcal{U}(-\frac{1}{2},\frac{1}{2}))(\tilde{\mathbf{c}}_{t,i}^a),
	\end{aligned}
\end{equation}
\begin{equation}
\begin{aligned}
	&P_{\tilde{\mathbf{c}}_{t}^{na}|\mathbf{m}_{t},\tilde{\mathbf{z}}_t^c,\tilde{\mathbf{c}}_{t}^a}(\tilde{\mathbf{c}}_{t}^{na}|\mathbf{m}_{t},\tilde{\mathbf{z}}_t^c,\tilde{\mathbf{c}}_{t}^a)\\&=\prod_{i}P_{\tilde{\mathbf{c}}_{t,i}^{na}|\mathbf{m}_{t,i},\tilde{\mathbf{z}}_t^c,\tilde{\mathbf{c}}_{t,<i},\tilde{\mathbf{c}}_{t,i}^a}(\tilde{\mathbf{c}}_{t,i}^{na}|\mathbf{m}_{t,i},\tilde{\mathbf{z}}_t^c,\tilde{\mathbf{c}}_{t,<i},\tilde{\mathbf{c}}_{t,i}^a)\\&=\prod_{i}(\mathcal{L}(\tilde{\mu}_{t,i}^{c,na},\tilde{\sigma}_{t,i}^{c,na})*\mathcal{U}(-\frac{1}{2},\frac{1}{2}))(\tilde{\mathbf{c}}_{t,i}^{na}),
\end{aligned}
\end{equation}
where $\tilde{\mathbf{z}}_t^c=\mathbf{z}_t^c+\mathbf{o}_t$ is the uniformly-noised hyperprior parameter.

The aforementioned learned mean and variance values are given as
\begin{equation}
\begin{aligned}
	(\tilde{\mu}_{t,i}^{c,a},\tilde{\sigma}_{t,i}^{c,a})&=g_{ep}(\mathbf{\varphi}_m^i,\mathbf{\varphi}_{ch}^i,\mathbf{\varphi}_z)\\&=g_{ep}(g_m(\mathbf{\mathbf{m}}_{t,\le i}),g_{ch}(\mathbf{\tilde{c}}_{t,\le i}),g_z(\mathbf{\tilde{z}}_t)),
\end{aligned}
\end{equation}
\begin{equation}
\begin{aligned}
	(\tilde{\mu}_{t,i}^{c,na},\tilde{\sigma}_{t,i}^{c,na})&=g_{ep}(\mathbf{\varphi}_m^i,\mathbf{\varphi}_{ch}^i,\mathbf{\varphi}_z,\mathbf{\varphi}_{lc}^i)\\&=g_{ep}(g_m(\mathbf{\mathbf{m}}_{t,\le i}),g_{ch}(\mathbf{\tilde{c}}_{t,\le i}),g_z(\mathbf{\tilde{z}}_t),g_{lc}(\mathbf{\tilde{c}}_{t,i}^a)),
\end{aligned}
\end{equation}
where $g_m(\cdot)$, $g_{ch}(\cdot)$, $g_z(\cdot)$, and $g_{lc}(\cdot)$ are the corresponding reference generators, $\mathbf{\varphi}_m^i$, $\mathbf{\varphi}_{ch}^i$, $\mathbf{\varphi}_z$, and $\mathbf{\varphi}_{lc}^i$ are the wireless channel reference, previously context reference, hyperprior reference, and local reference for entropy coding.

After that, since we have no prior beliefs about $\tilde{\mathbf{z}}_{t}^c$, non-parametric fully factorized density is utilized to model the hyperprior distribution as
\begin{align}
	P_{\tilde{\mathbf{z}}_{t}^c}(\tilde{\mathbf{z}}_{t}^c)=\prod_{j}(P_{\mathbf{z}_{t,j}^c|\psi^{(j)}}(\mathbf{z}_{t,j}^c|\psi^{(j)})*\mathcal{U}(-\frac{1}{2},\frac{1}{2}))(\tilde{\mathbf{z}}_{t,j}^c),
\end{align}
where $\psi^{(j)}$ encapsulates all the parameters of $P_{\mathbf{z}_{t,j}^c|\psi^{(j)}}$.

With the learned entropy model, the allocated channel bandwidth cost for the feature context is formulated as
\begin{equation}
\begin{aligned}
	k_{t,i}^c&=\eta_{t,i}^cr_{t,i}^c\\&=-\eta_{t,i}^c(\log P_{\bar{\mathbf{c}}_{t,i}^{a}|\mathbf{m}_{t,\le i},\bar{\mathbf{z}}_t^c,\bar{\mathbf{c}}_{t,<i}}(\bar{\mathbf{c}}_{t,i}^{a}|\mathbf{m}_{t,\le i},\bar{\mathbf{z}}_t^c,\bar{\mathbf{c}}_{t,<i})\\&+\log P_{\bar{\mathbf{c}}_{t,i}^{na}|\mathbf{m}_{t,\le i},\bar{\mathbf{z}}_t^c,\bar{\mathbf{c}}_{t,<i},\bar{\mathbf{c}}_{t,i}^a}(\bar{\mathbf{c}}_{t,i}^{na}|\mathbf{m}_{t,\le i},\bar{\mathbf{z}}_t^c,\bar{\mathbf{c}}_{t,<i},\bar{\mathbf{c}}_{t,i}^a)),
\end{aligned}
\end{equation}
where $\eta_{t,i}^c$ refers to the $i$-th group adjust hyperparameter of feature context.

Unlike \cite{dvst}, where $\eta_{t,i}^c$ remains unchanged throughout the entropy coding computation stage, CVST requires distinct $\eta_{t,i}^c$ values for different channel conditions and varying levels of semantic importance across feature groups due to its group-level feature optimization. Within the context-channel correlation map, each column is normalized to the sum 1, representing the weight of each feature group within a specific subchannel. Optimization of matching context-channel pairs that maximizes the sum of these weights is a typical linear assignment problem, which can be properly solved by Hungarian algorithm. The problem for such context-channel pair assignment is formulated within each gather as
\begin{equation}
\begin{aligned}
	& \max \sum_{i}\sum_{j}m_{t,ij}* s_{t,ij}\\
	& s.t. \left\{\begin{matrix}\sum_{i} s_{t,ij}=1
		\\
		\sum_{j} s_{t,ij}=1
	\end{matrix}\right.
\end{aligned}
\end{equation}
where $s_{t,ij}$ refers to the linear assignment decision of each context-channel pair.

Follow this way, each $\eta_{t,i}^c$ value can be computed as
\begin{align}
	\eta_{t,i}^c=\eta_{d}^c*\mathbf{m}_{t,\mathrm{Hun}(\mathbf{m}_t)}*N_t,
\end{align}
where $\mathrm{Hun}(\cdot)$ is the classical Hungarian algorithm for solving such linear assignment problem with proper indexes as output. $\mathbf{m}_{t,\mathrm{Hun}(\mathbf{m}_t)}$ is thus the weight of selected index in the optimal path with a vector form. $\eta_d^c$ is the default hyperparameter for rate adjustment. $N_t$ scales the $\eta_{t,i}^c$ value.

Then, the total channel bandwidth cost is collected as
\begin{equation}
\begin{aligned}
	k_{t}^c&=-\sum_{i}\eta_{t,i}^c(\log P_{\bar{\mathbf{c}}_{t,i}^{a}|\mathbf{m}_{t,\le i},\bar{\mathbf{z}}_t^c,\bar{\mathbf{c}}_{t,<i}}(\bar{\mathbf{c}}_{t,i}^{a}|\mathbf{m}_{t,\le i},\bar{\mathbf{z}}_t^c,\bar{\mathbf{c}}_{t,<i})\\&+\log P_{\bar{\mathbf{c}}_{t,i}^{na}|\mathbf{m}_{t,\le i},\bar{\mathbf{z}}_t^c,\bar{\mathbf{c}}_{t,<i},\bar{\mathbf{c}}_{t,i}^a}(\bar{\mathbf{c}}_{t,i}^{na}|\mathbf{m}_{t,\le i},\bar{\mathbf{z}}_t^c,\bar{\mathbf{c}}_{t,<i},\bar{\mathbf{c}}_{t,i}^a)).
\end{aligned}
\end{equation}

Similarly, the total allocated channel bandwidth cost for the motion vector is formulated as
\begin{equation}
	\begin{aligned}
	k_{t}^v&=-\sum_{i}\eta_{t,i}^v(\log P_{\bar{\mathbf{v}}_{t,i}^{a}|\mathbf{m}_{t,\le i},\bar{\mathbf{z}}_t^v,\bar{\mathbf{v}}_{t,<i}}(\bar{\mathbf{v}}_{t,i}^{a}|\mathbf{m}_{t,\le i},\bar{\mathbf{z}}_t^v,\bar{\mathbf{v}}_{t,<i})\\&+\log P_{\bar{\mathbf{v}}_{t,i}^{na}|\mathbf{m}_{t,\le i},\bar{\mathbf{z}}_t^v,\bar{\mathbf{v}}_{t,<i},\bar{\mathbf{v}}_{t,i}^a}(\bar{\mathbf{v}}_{t,i}^{na}|\mathbf{m}_{t,\le i},\bar{\mathbf{z}}_t^v,\bar{\mathbf{v}}_{t,<i},\bar{\mathbf{v}}_{t,i}^a)),
	\end{aligned}
\end{equation}
where $\bar{\mathbf{z}}_t^v= \lfloor \mathbf{z}_t^v\rceil=\lfloor h_r^v(\mathbf{v}_t)\rceil$, $\eta_{t,i}^v$ is the rate adjust hyperparameter of motion vector.

Finally, the transmission cost for the CVST is formulated as
\begin{align}
	k_{t}=k_{t}^c + k_{t}^v + k_{t}^{c_z} + k_{t}^{v_z},
\end{align}
where $k_{t}^{c_z}$ and $k_{t}^{v_z}$ refer to the hyperprior vector transmission bandwidth cost of feature context and motion vector, respectively. It can be learned according to Eq. (12).

The illustration of proposed MR-VLRC is shown in Fig. \ref{fig_3}. Since the hyperprior entropy model enables flexible rate adjustment for transmitted semantic codewords, multiple references strength the rational allocation in consideration of spatial, temporal, and channel characteristics. For each feature context group, the original context is first split into anchored and non-anchored parts. The entropy of the anchored part is first learned with the help of diverse references $\mathbf{\varphi}_m^i$, $\mathbf{\varphi}_{ch}^i$, and $\mathbf{\varphi}_z$. Then, the current entropy of the anchored part is collected as $\mathbf{\varphi}_{lc}^i$ for the non-anchored part. The complete entropy of feature context is aggregated by both anchored part and non-anchored part. Finally, the hyperparameter $\eta^c$ for rate adjustment is learned through the context-channel correlation map.

\begin{figure*}[htbp]
	\centering
	\includegraphics[width=7.0in]{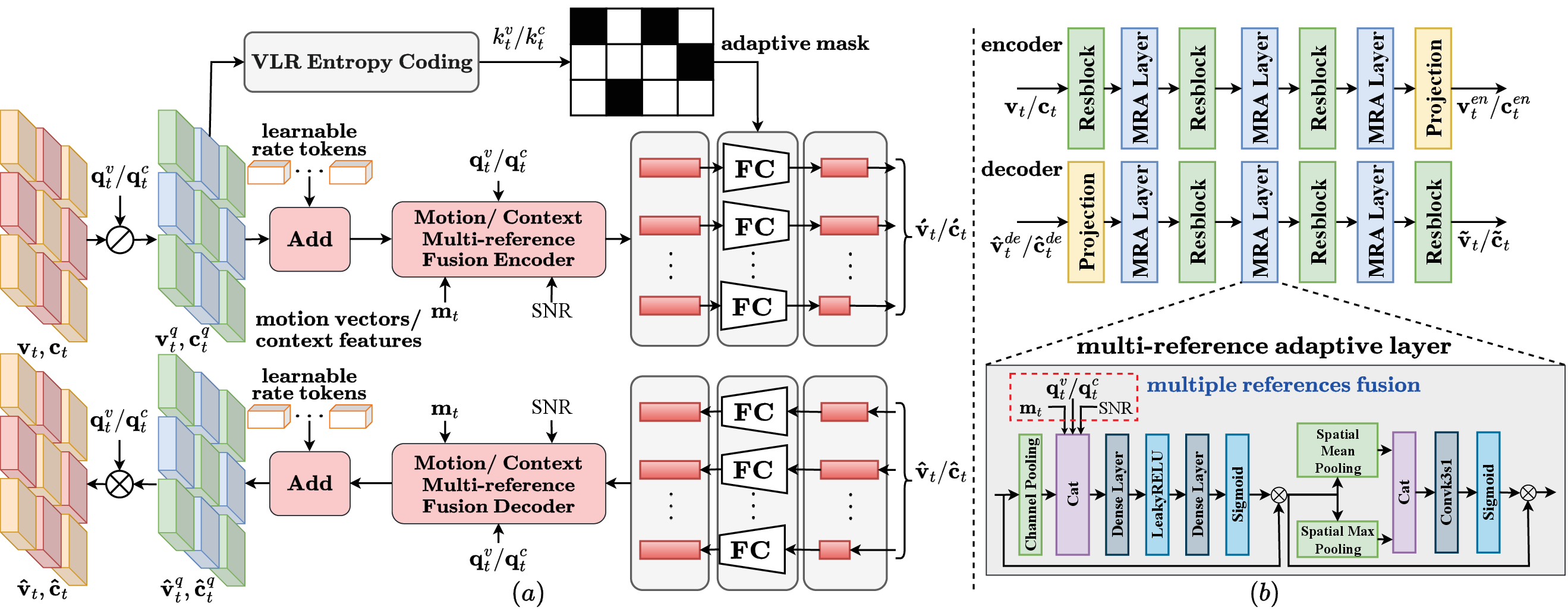}
	\caption{(a) The structure of motion vector/context VLRC coder including checkerboard feature modulation, VLR entropy coding, multi-references fusion coder, and rate adaptation coder. (b) The structure of multi-reference fusion motion vector/context coder.}
	\label{fig_4}
\end{figure*}

\subsubsection{Variable Rate Coding for Wireless Video Transmission}
Building upon the multi-reference variable length coding framework, we further extend CVST to support variable rate coding. Crucially, this enables CVST to evaluate multiple rate points using a single model trained only once. To achieve this, we adopt a feature modulation scheme inspired by \cite{dcvc-hem}, analogous to adapting the quantization parameter in traditional video coding for different quantization levels. Within the proposed CVST framework, we integrate this feature modulation mechanism into our multi-reference variable length entropy coding model. This integration provides fine-grained feature modulation applicable to both anchored and non-anchored feature components.

Taking the feature context for instance, the feature modulation term $\mathbf{q}_{t}^{c}$ is given as  
\begin{align}
	\mathbf{q}_{t}^{c} & = \begin{cases}\mathbf{q}_{t}^{c,a} & = r_{\mathrm{global},\lambda}^c\cdot\mathbf{r}_{t}^{c,a},
		\\
		\mathbf{q}_{t}^{c,na} & = r_{\mathrm{global},\lambda}^c\cdot\mathbf{r}_{t}^{c,na},
	\end{cases}
\end{align}
where $\mathbf{q}_{t}^{c,a}$ and $\mathbf{q}_{t}^{c,na}$ refer to the archored and non-anchored feature modulation terms for the feature context. $r_{\mathrm{global},\lambda}^c$ is the global feature context modulation value which highly reflects different $\lambda$ values during training. It works as the quantization parameter in traditional video coding schemes, e.g. VVC, as smaller global modulation values refer to better reconstructed video qualities. $\mathbf{r}_{t}^{c,a}\in\mathbb{R}^{L}$ and $\mathbf{r}_{t}^{c,na}\in\mathbb{R}^{L}$ refer to the feature channel dimension modulation term which are presented as fine-grained modulation for the transmitted feature context.

Similarly, the feature modulation term $\mathbf{q}_{t}^{v}$ for the motion vector is given as  
\begin{align}
	\mathbf{q}_{t}^{v} & = \begin{cases}\mathbf{q}_{t}^{v,a} & = r_{\mathrm{global},\lambda}^v\cdot\mathbf{r}_{t}^{v,a},
		\\
		\mathbf{q}_{t}^{v,na} & = r_{\mathrm{global},\lambda}^v\cdot\mathbf{r}_{t}^{v,na},
	\end{cases}
\end{align}

The combination of global term and channel-wise terms jointly promotes the coarse-to-fine feature modulation.

With $\mathbf{q}_{t}^{c}$ and $\mathbf{q}_{t}^{v}$, the entropy coding can be modulated for both anchored and non-anchored parts as
\begin{align}
\begin{Bmatrix}
	\mathbf{v}_t^q\\\mathbf{c}_t^q
\end{Bmatrix}=\begin{Bmatrix}
	\mathbf{v}_t\oslash\mathbf{q}_{t}^{v}\\\mathbf{c}_t\oslash\mathbf{q}_{t}^{c}
\end{Bmatrix}=\begin{Bmatrix}
	\mathrm{Mer}(\mathbf{v}_t^a\oslash\mathbf{q}_{t}^{v,a},\mathbf{v}_t^{na}\oslash\mathbf{q}_{t}^{v,na})\\\mathrm{Mer}(\mathbf{c}_t^a\oslash\mathbf{q}_{t}^{c,a},\mathbf{c}_t^{na}\oslash\mathbf{q}_{t}^{c,na})
\end{Bmatrix},
\end{align}
where $\mathrm{Mer}(\cdot)$ refers to the merge operation for combining the modulated anchored and non-anchored parts together.

At the decoder side, the received codewords are demodulated as
\begin{align}
\begin{Bmatrix}
	\mathbf{\tilde{v}}_t^q\\\mathbf{\tilde{c}}_t^q
\end{Bmatrix}=\begin{Bmatrix}
	\mathbf{\hat{v}}_t\otimes\mathbf{q}_{t}^{v}\\\mathbf{\hat{c}}_t\otimes\mathbf{q}_{t}^{c}
\end{Bmatrix}=\begin{Bmatrix}
	\mathrm{Mer}(\mathbf{\hat{v}}_t^a\otimes\mathbf{q}_{t}^{v,a},\mathbf{\hat{v}}_t^{na}\otimes\mathbf{q}_{t}^{v,na})\\\mathrm{Mer}(\mathbf{\hat{c}}_t^a\otimes\mathbf{q}_{t}^{c,a},\mathbf{\hat{c}}_t^{na}\otimes\mathbf{q}_{t}^{c,na})
\end{Bmatrix}.
\end{align}

In this way, rather than directly modulate the whole feature in \cite{dcvc-hem}, the checkerboard-typed feature modulation terms enable the subtle rate adaptivity based on both spatial locality and feature channel importance.

\subsubsection{Correlation between feature context transmission and motion vector transmission}

While the proposed efficient MR-VLRC enables CVST to achieve variable constant CBR conditions through VLC, a limitation exists: feature context and motion vector undergo independent entropy encoding without exploiting their inherent relationship. Crucially, in context-based wireless video transmission, motion vectors exhibit significantly higher compressibility than feature context. Consequently, motion vectors require fewer channel bandwidth resources than the feature context. However, \cite{dvst} allocates identical entropy coding priority to both components via the same $\eta$ value, which neglects their differential compression characteristics and suggests clear room for improvement.

Based on the multi-reference entropy coding for VLC, the $\eta$ values are changeable according to the context-channel correlation map to adapt to variable channel states. Furthermore, we constrain the feature context $\eta^{c}$ values and motion vector $\eta^{v}$ values with separately predefined value range to achieve a balance between feature context and motion vector transmission. The value constraint is formulated as
\begin{align}
	\tilde{\eta}^c=\mathrm{Cla}(\eta^c,[\eta^c_{\mathrm{min}},\eta^c_{\mathrm{max}}]),
\end{align}
\begin{align}
	\tilde{\eta}^v=\mathrm{Cla}(\eta^v,[\eta^v_{\mathrm{min}},\eta^v_{\mathrm{max}}]),
\end{align}
where $\tilde{\eta}^c$ and $\tilde{\eta}^v$ are the constrained $\eta$ values for the feature context and motion vector to substitute ${\eta}^c$ and ${\eta}^v$, respectively. $[\eta^c_{\mathrm{min}},\eta^c_{\mathrm{max}}]$ and $[\eta^v_{\mathrm{min}},\eta^v_{\mathrm{max}}]$ are the corresponding lower bound and upper bound, $\mathrm{Cla}(\cdot,[\cdot, \cdot])$ refers to the cut off operation for regulating $\eta$ value to a predefined range.

In this way, the $\eta$ values are able to be restricted to a specific value range. In practice, $\tilde{\eta}^c$ usually has higher value than $\tilde{\eta}^v$, thus enabling the unequal channel bandwidth allocation between feature context and motion vector.

\subsection{Multi-reference Fusion for Robust Coding}
The structure of the motion vector/context VLRC coder is depicted in Fig. \ref{fig_4}. Take the feature context coding as an example, the concatenated and refined feature context $\mathbf{c}_t$ is first modulated by the $\mathbf{q}_t^c$ as $\mathbf{c}_t^q$. Subsequently, VLR entropy coding generates an adaptive mask and learnable rate tokens to produce the rate-aware feature context. After that, context reference fusion encoder encodes $\mathbf{c}_t$ by incorporating wireless channel references, e.g. current SNR and context-channel correlation map $\mathbf{m}_t$, and rate-adaptive reference $\mathbf{q}_t^c$ to perform robust coding. The structure of multi-reference fusion coder is shown in Fig. \ref{fig_4}(b). The encoder employs a series of Resnetblock and multi-reference adaptive (MRA) Layer to fuse the channel-aware and rate-aware references into the transmitted feature context. The MRA Layer is divided into two paths. One is the original feature context. The other is the reference-aware modulation term. It utilizes channel pooling to transform the feature context into the feature channel dimension and then concatenate with the expanded SNR, $\mathbf{m}_t$, and $\mathbf{q}_t^c$. Through several linear and activation layers, the modulation term is multiplied with the feature context. Then the spatial mean pooling along with spatial max pooling formulates an extra modulation term to further tune the modulated feature context into $\mathbf{c}_t^{en}$. Finally, using this rate-aware and CSI-aware $\mathbf{c}_t^{en}$, the dynamic rate-adaptive network from \cite{NTSCC} learns variable length transmitted codewords $\mathbf{\acute{c}}_t$, while the decoder performs inverse operations relative to the encoder.

\section{Deployment of CVST Framework}
In this section, we provide the deployment details of CVST framework, including network structure, training loss, and training strategy, respectively.
\subsection{Network Structure}
For the network structure, we consider the lightweight and concise CNNs as the main network backbones for the overall designs.
\begin{figure}[htbp]
	\centering
	\includegraphics[width=3.4in]{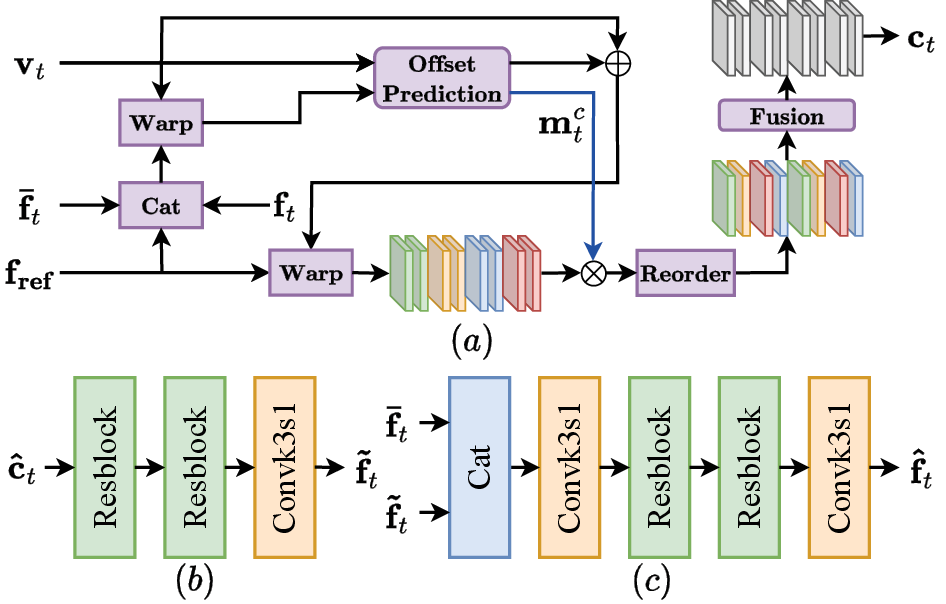}
	\caption{The network structures of different modules. (a) Context generator; (b) Frame regeneration; (c) Frame refinement.}
	\label{fig_5}
\end{figure}
$\textbf{Semantic Encoder \& Decoder:}$ For the semantic coder, we adapt the lightweight CNN-based Resnet strucuture as the network backbone, following the work \cite{mmvc}.

$\textbf{Motion Estimation \& Compensation:}$ For the motion estimation $\&$ compensation module, we adapt the same network structure as \cite{wvsc} for producing offset and predicted frames.

$\textbf{Context Generation:}$ For the context generation module, the structure is shown in Fig. \ref{fig_5}(a). Following \cite{dcvc-dc}, the context generation module contains spatial-related input $\mathbf{\bar{f}}_t$, $\mathbf{f}_t$ along with temporal-related input $\mathbf{v}_t$, $\mathbf{f}_\mathbf{ref}$. The offset prediction part estimates offset and mask for the feature fusion. It is notable that fused features are reordered to perform cross group fusion. In this way, the generated context $\mathbf{c}_t$ has the global feature characteristic of the current frame.

$\textbf{Frame Regeneration:}$ For the frame regeneration module, as shown in Fig. \ref{fig_5}(b), it is also a CNN-based structure.

$\textbf{Frame Refinement:}$ For the frame refinement module, the network structure is shown in Fig. \ref{fig_5}(c).

\begin{figure*}[htbp]
	\centering  
	\subfigure[UVG (CBR = 0.027)]{
		\includegraphics[width=0.32\linewidth]{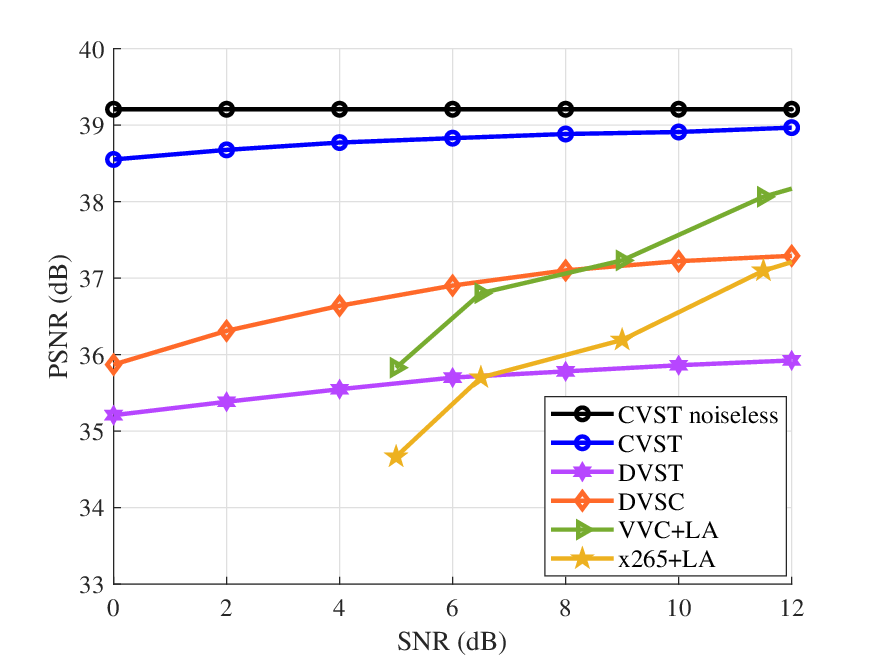}}
	\subfigure[HEVC ClassA (CBR = 0.040)]{
		\includegraphics[width=0.32\linewidth]{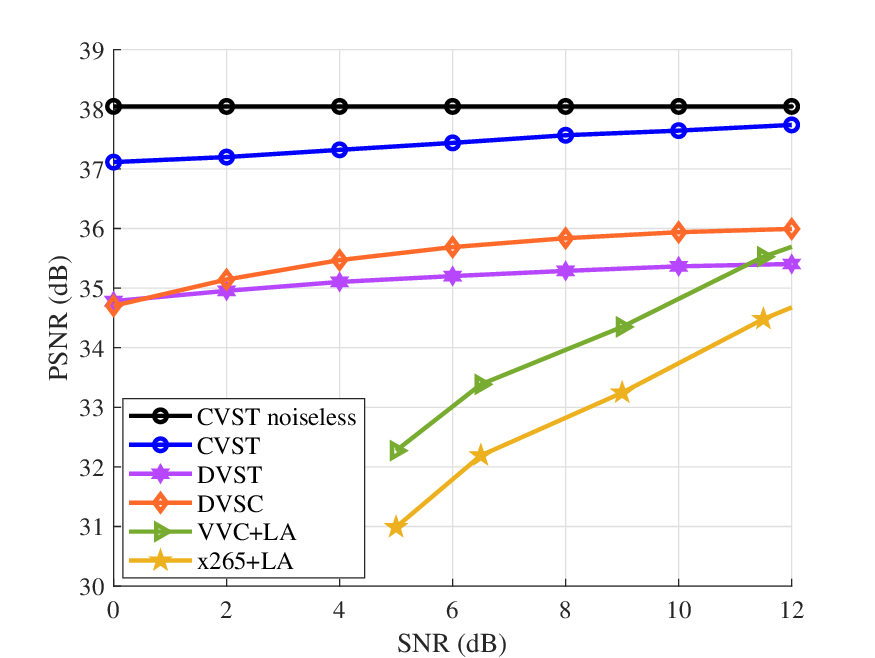}}
	\subfigure[HEVC ClassB (CBR = 0.043)]{
		\includegraphics[width=0.32\linewidth]{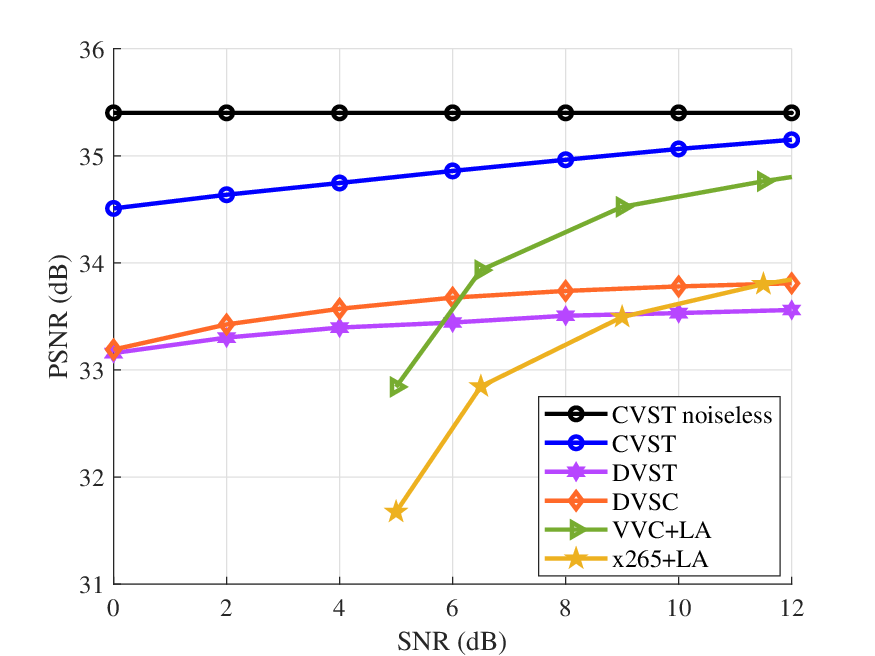}}
	\subfigure[HEVC ClassC (CBR = 0.054)]{
		\includegraphics[width=0.32\linewidth]{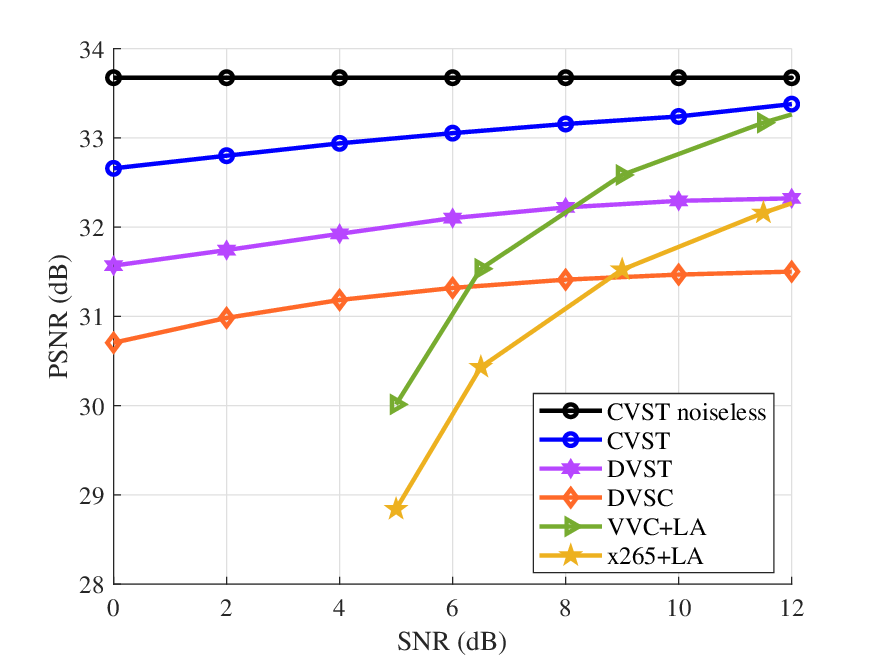}}
	\subfigure[HEVC ClassD (CBR = 0.064)]{
		\includegraphics[width=0.32\linewidth]{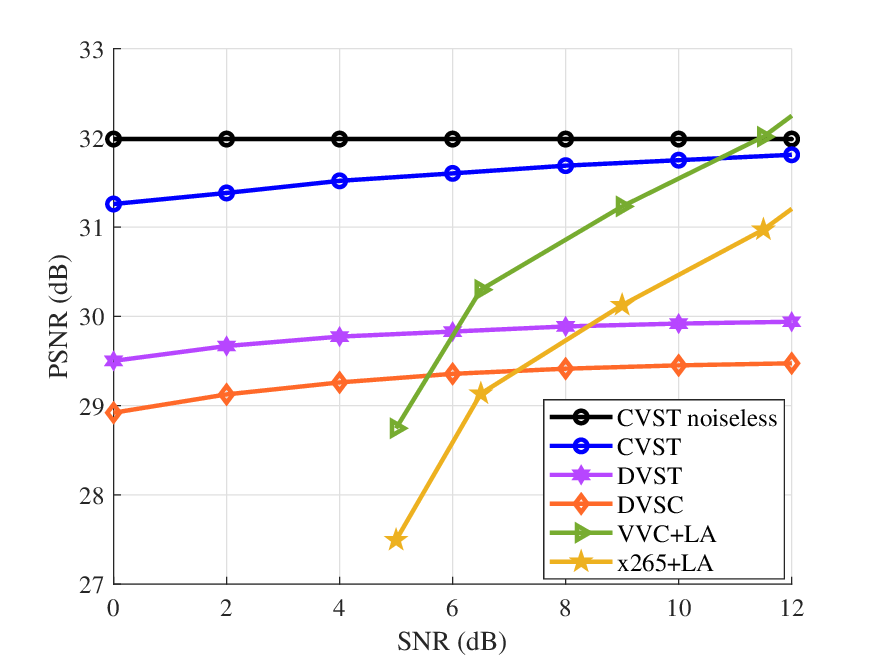}}
	\caption{Quality of the reconstructed video frames versus the SNRs under MIMO fading channels.}
	\label{fig_6}
\end{figure*}

\subsection{Training Loss and Strategy}
Since CVST has many modules to train, it is hard to directly train all the modules together in an efficient manner. Thus, the multi-stage training schemes seem to be a good choice. The CVST is trained in a progressive manner with several stages.

We first train the CVST without VLRC-related modules to obtain a reasonably satisfying initialized result for CVST. The training loss is simply given as
\begin{align}
	L_{t}=D_t(\mathbf{x}_t,\mathbf{\hat{x}}_t),
\end{align}
where $D_t(\cdot,\cdot)$ is the distortion for the frame reconstruction loss, which can be set as mean square error (MSE) in default or other perceptual losses.

Then, we introduce the VLRC related modules to jointly train the CVST. The optimization target is given as the rate distortion loss as
\begin{equation}
\begin{aligned}
	L_{t}&=k_t+\lambda\cdot(D_t+D^{\mathrm{ntc}}_t)\\&=k_t^{c}+k_t^{c_z}+k_t^{v}+k_t^{v_z}+\lambda\cdot(D_t(\mathbf{x}_t,\mathbf{\hat{x}}_t)+D^{\mathrm{ntc}}_t(\mathbf{x}_t,\mathbf{\bar{x}}_t)),
\end{aligned}
\end{equation}
where $\lambda$ is the randomly selected Lagrange multiplication for the rate distortion trade off from a set of predefined $\lambda$ values. The multiple $\lambda$ values training not only enables the rate adaptation optimization for the specific rate point but also determines the feature modulation term for variable CBRs. $D^{\mathrm{ntc}}_t(\cdot,\cdot)$ is the distortion loss between the original frames and quantized reconstructed frames $\mathbf{\bar{x}}_t$ by variable length coding, which performs as a reweighting term for keeping the variable length and rate training stable. 

With the above progressive training strategy, the channel-adaptive and rate-adaptive performances can be gradually achieved.

\section{Numerical Results}
In this section, we present numerical results to evaluate the effectiveness of proposed CVST for wireless video transmission.

\subsection{Experimental Setups}
\subsubsection{Datasets}

For the wireless video semantic transmission, we quantify the performances of proposed CVST versus other benchmarks over the Vimeo-90k dataset \cite{vimeo}, which is a widely-used dataset consisting of 89800 various video correlations for deep video compression. The training dataset is split according to \cite{dv-c}. During model training, images are randomly cropped to 256$\times$256$\times$3. While for model testing, we adapt UVG dataset \cite{uvg} (1920 × 1080) and HEVC test datasets \cite{hevc} including ClassA (2560 × 1600), ClassB (1920 × 1080), ClassC (832 × 480), and ClassD (416 × 240) for evaluating the effectiveness of proposed CVST over various contents, resolutions, and frame rates. For the wireless channels, unless specifically mentioned, we choose the 8$\times$8 MIMO Rayleigh fading channels for both training and testing similar to the single-user MIMO environment deployment in \cite{semimo}.  

\subsubsection{Model Deployment Details}
The channel dimension $L$ is set as 64, for both motion vectors and context. Since CVST mainly focuses on the robust and flexible P frame transmission, The I frame transmission network follows the VLC fixed-rate wireless image transmission and adapts the semantic coder and multi-reference fusion coder of CVST structures. SNR and $\lambda$ values are randomly sampled for each video frame batch during training stage to acquire SNR-adaptive and CBR-adaptive results in various testing conditions. The SNR set is defined as [0, 2, 4, 6, 8, 10, 12, 14] dB while $\lambda$ set is defined as [0.015, 0.06, 0.12, 0.20, 0.32]. For model training, we use variable learning rate, which decreases step-by-step from 5e-5 to 2e-5 along with different epochs. The batch size is set as 16. The training GoP size for the Vimeo-90k is 7 while testing GoP size is 12 to evaluate the temporal consistency. The whole framework is optimized with AdamW algorithm. Under the $8\times8$ MIMO fading channel conditions, DL-based schemes employ the SVD algorithm for precoding and detection. To speed up the training process, automatic mixed precision with BFloat16 is adopted during training and testing. All the experiments are run in RTX 4090 GPUs with Pytorch2.0.

\subsubsection{Comparison Benchmarks}
In the experiments, several benchmarks are given as below:

$\textbf{DVSC}$: The DL-empowered deep video transmission framework \cite{dvsc} with SNR-adaptive channel coder and semantic restoration at the receiving end.

$\textbf{DVST}$: The wireless video semantic transmission framework \cite{dvst} adapting contexual coding and rate-adaptive transmission.

$\textbf{CVST noiseless}$: The ideal noiseless transmission case of CVST. The context-channel correlation map generation and wireless channel-related embeddings e.g., $\mathbf{m}_t$ and SNR, are excluded. 

$\textbf{VVC/x265+LDPC+QAM+RI+WF}$: The SSCC scheme with VVC/H.265 video codec as source coding and LDPC as channel coding, along with the quadrature amplitude modulation (QAM). For the wireless MIMO transmission and error detection, SVD precoding, random interleave (RI) method and waterfilling (WF) power allocation are also adapted. 

$\textbf{VVC/x265+LA}$: Based on the `VVC/x265+LDPC+QAM+ RI+WF', link adaptation (LA) is adapted for adjusting LDPC code rate and QAM modulation order according to SNRs.

Note that DVSC and DVST are the DL-based wireless video transmission schemes, whose parameter configurations are presented in Tab. \ref{table2}. Traditional VVC/H.265+LDPC+QAM +RI+WF are existing separated coding schemes. The video codec VVC is adopted by the vvenc \cite{vvenc} while H.265 is adopted by the x265 in ffmpeg-python \cite{ffmpeg}, which balance the coding efficiency and performance in the practical deployment. Both VVC and H.265 adopt the low-delay mode and GoP size 32. The 5G NR LDPC \cite{sionna} along with random interleave is adapted for channel coding.

\begin{table*}[htbp]
	\centering
	\caption{Configurations of benchmarks. CNN($k$,$p$,$s$) is the CNN network with kernel size $k$, padding $p$, stride $s$. dim is the hidden dimension, which is set aligned with proposed CVST. Res is the residual block network. (I)GDN is the (inverse) generalized divisive normalization.}
	\label{table2}
	
	\begin{tabular}{|c|c|c|c|}  
		\hline 
		& & &\\[-6pt] 
		Framework&Module&Backbone&Deployment\\
		\hline
		& & &\\[-6pt]  
		\multirow{5}{*}{DVST}&Contextual Analysis Transform&[CNN(3,1,1)+GDN+Res]$\times$N+CNN(3,1,1)&N=3,dim=64\\ 
		\cline{2-4}
		& & &\\[-6pt]  
		&Contextual Synthesis Transform&[CNN(3,1,1)+IGDN+Res]$\times$N+CNN(3,1,1)&N=3,dim=64\\
		\cline{2-4}
		& & &\\[-6pt]  
		&Contextual Deep JSCC Encoder/ Decoder&Swin Transformer \cite{swin}&l=4,h=8,dim=64\\
		\cline{2-4}
		& & &\\[-6pt]  
		&Motion Estimation&optic flow \cite{optic}&dim=64\\
		\cline{2-4}
		& & &\\[-6pt]  
		&Entropy Model $\&$ Rate Allocation&hyperprior entropy coding \cite{hpc}&dim=64\\
		\hline
		& & &\\[-6pt] 
		\multirow{5}{*}{DVSC}&Semantic Encoder&[CNN(3,1,1)+GDN]$\times$N+CNN(3,1,1)&N=3,dim=64\\
		\cline{2-4}
		& & &\\[-6pt] 
		&Semantic Decoder&DCNN(3,1,1)+[IGDN+DCNN(3,1,1)]$\times$N&N=3,dim=64\\
		\cline{2-4}
		& & &\\[-6pt] 
		&Channel Encoder&[Res+SNR adapt layer]$\times$N+CNN(3,1,1)&N=2,dim=64\\
		\cline{2-4}
		& & &\\[-6pt] 
		&Channel Decoder&DCNN(3,1,1)+[Res+SNR adapt layer]$\times$N&N=2,dim=64\\
		\cline{2-4}
		& & &\\[-6pt] 
		&Semantic Correction&U-Net \cite{unet}+RCAB \cite{rcab}&dim=64\\
		\hline
	\end{tabular}
\end{table*}

\subsubsection{Evaluation Metrics}

We leverage the widely used pixel-wise metric peak signal-to-noise ratio (PSNR), perceptual-level multi-scale structural similarity (MS-SSIM) and learned perceptual image patch similarity (LPIPS) as measurements for the reconstructed image quality. 

\subsection{Results Analysis}

\subsubsection{Performance for Different SNRs}
We first evaluate the anti-noise performances of CVST under MIMO fading channels with specific CBRs, where CBR refers to the bandwidth compression ratio between transmitted semantic codewords and original video signals. Here a single CVST model with perfect CSI is trained without variable rate coding modules for multi-SNR assessment. As shown in Fig. \ref{fig_6}, it is clearly to observe that CVST outperforms all other benchmarks. Compared to the DL-based wireless video transmission scheme, CVST consistently outperforms all benchmarks: it surpasses DVST and DVSC about 2 dB over UVG dataset, confirming that context video coding combined with multi-reference variable length coding significantly enhances transmission robustness. Moreover, CVST also exhibits markedly slower degradation than DVST and DVSC under different noise intensities, which means that CVST enables channel-adaptation against channel fading and noise throughout the whole framework. For the traditional separated coding schemes, we adapt the `x265+LA' and the `VVC+LA' as benchmarks, where LDPC code rate and QAM modulation order are set according to SNRs to achieve practical performances. Compared to traditional schemes, CVST provides much more performance gain and stability, avoiding the cliff effects seen in separated coding under harsh channels. It is also worth noting that CVST achieves relatively similar performances against the CVST noiseless, which employs no wireless channel noise during transmission. From Fig. \ref{fig_6}(a) to (e), the performance gap roughly widens with increasing frame dimensions, highlighting exceptional adaptability of CVST for high-resolution videos.

\subsubsection{Performance for Different CBRs}
\begin{figure*}[htbp]
	\centering  
	\subfigure[UVG]{
		\includegraphics[width=0.32\linewidth]{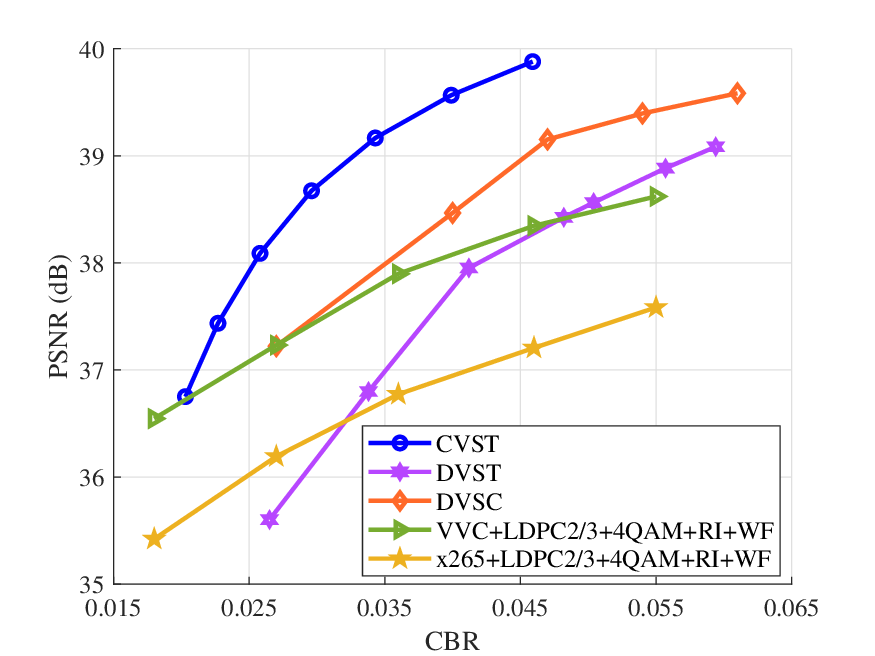}}
	\subfigure[HEVC ClassA]{
		\includegraphics[width=0.32\linewidth]{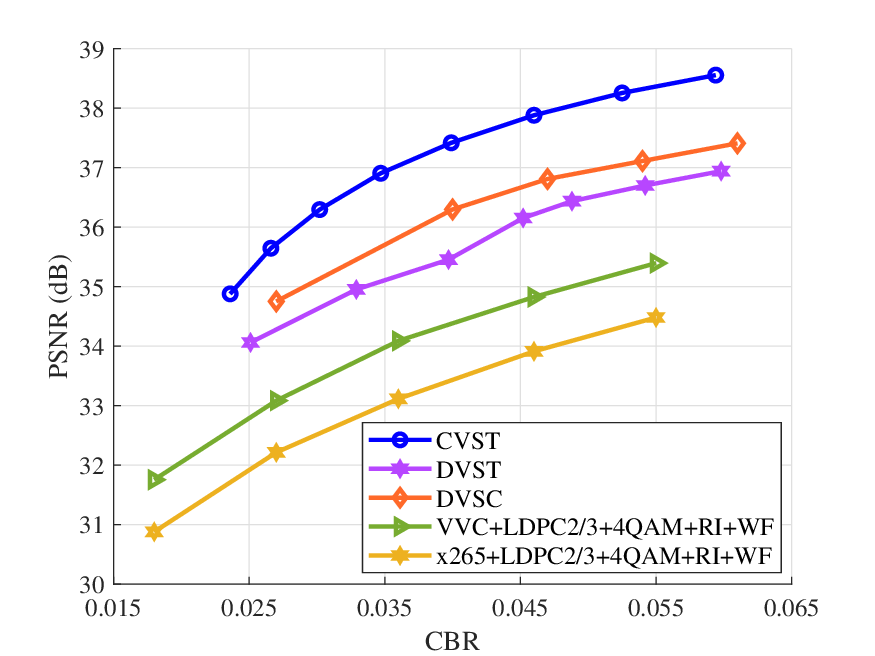}}
	\subfigure[HEVC ClassB]{
		\includegraphics[width=0.32\linewidth]{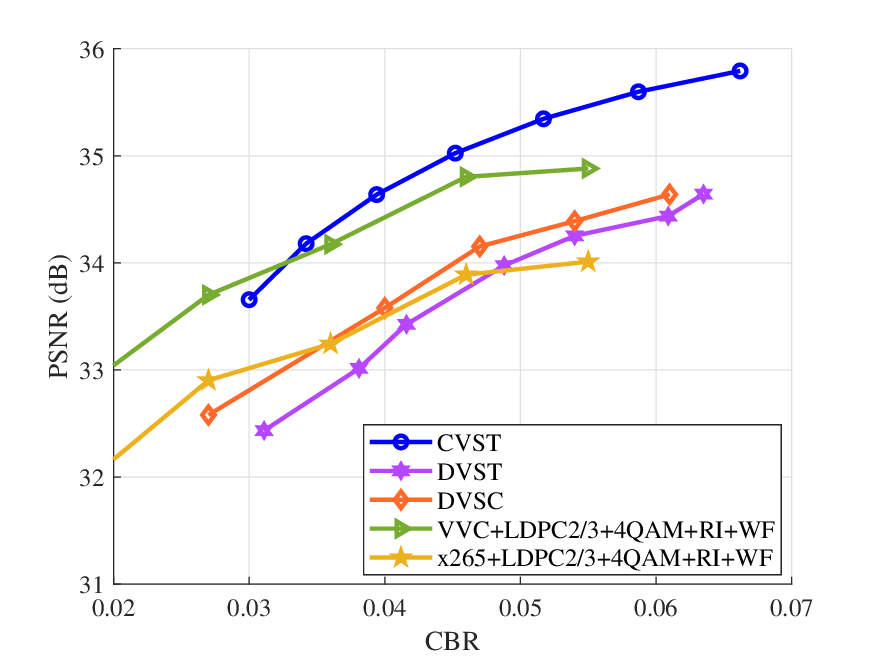}}
	\subfigure[HEVC ClassC]{
		\includegraphics[width=0.32\linewidth]{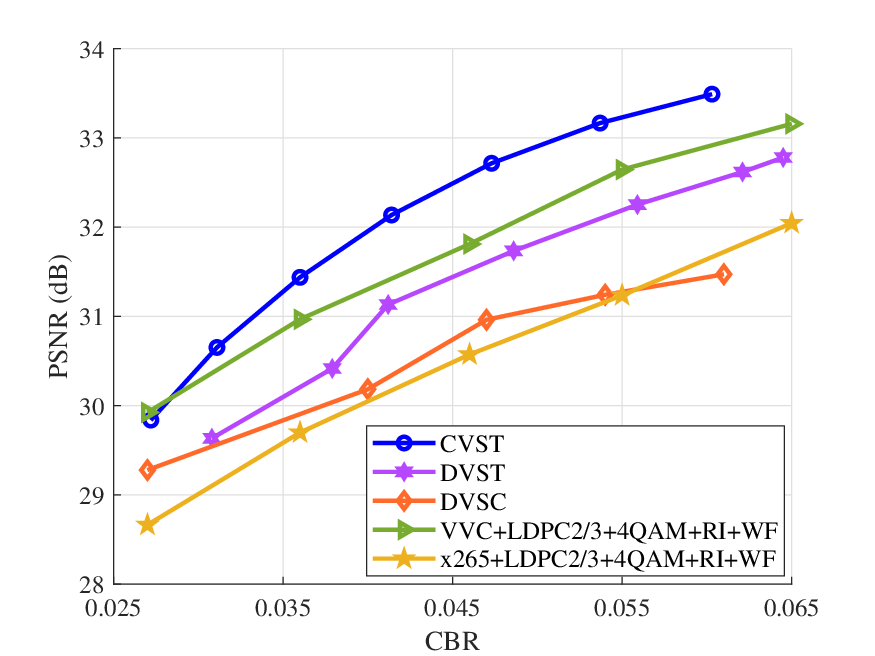}}
	\subfigure[HEVC ClassD]{
		\includegraphics[width=0.32\linewidth]{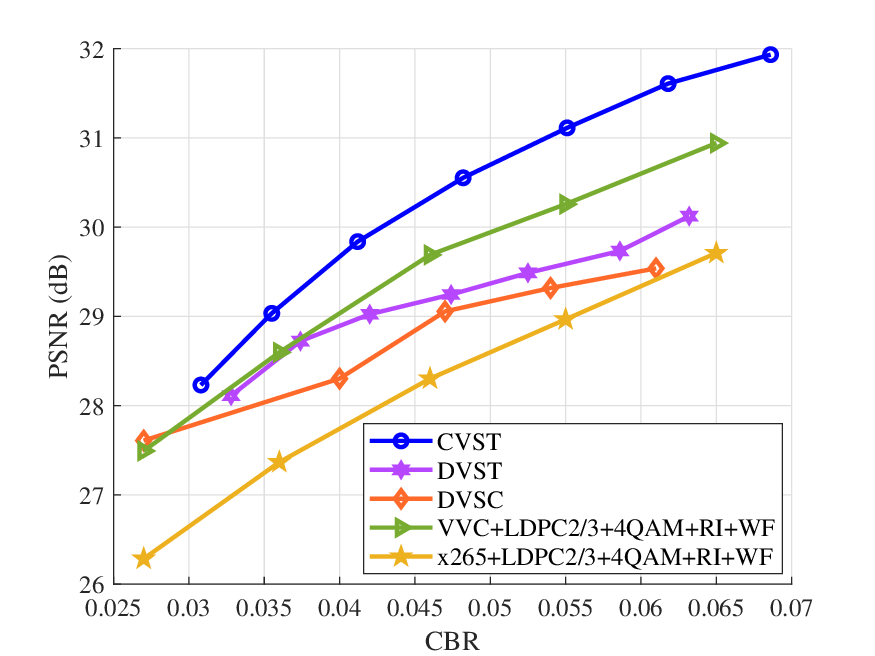}}
	\caption{Quality of the reconstructed video frames versus the CBRs under MIMO fading channels (SNR = 9).}
	\label{fig_7}
\end{figure*}

We next evaluate CVST's bandwidth compression performance under MIMO fading channels at SNR = 9 dB. As shown in Fig. \ref{fig_7}, CVST consistently achieves significant performance gains over all comparative schemes. Notably, the performance gaps between CVST and both DVST and DVSC widen as the CBR increases. This advantage stems primarily from the proposed MR-VLRC module whose well-designed feature modulation enables efficient variable length coding across diverse rate points within a single model, eliminating the need to train multiple models for different rates. Furthermore, CVST outperforms traditional VVC-based schemes (VVC+LDPC+QAM+RI+WF), demonstrating the superior compression efficiency of its jointly optimized context transmission and entropy coding. To conclude, CVST maintains stable performance gains across varying resolutions, video content types, and motion complexities, highlighting its flexibility for diverse video transmission scenarios.

\begin{figure}[htbp]
	\centering
	\includegraphics[width=3.6in]{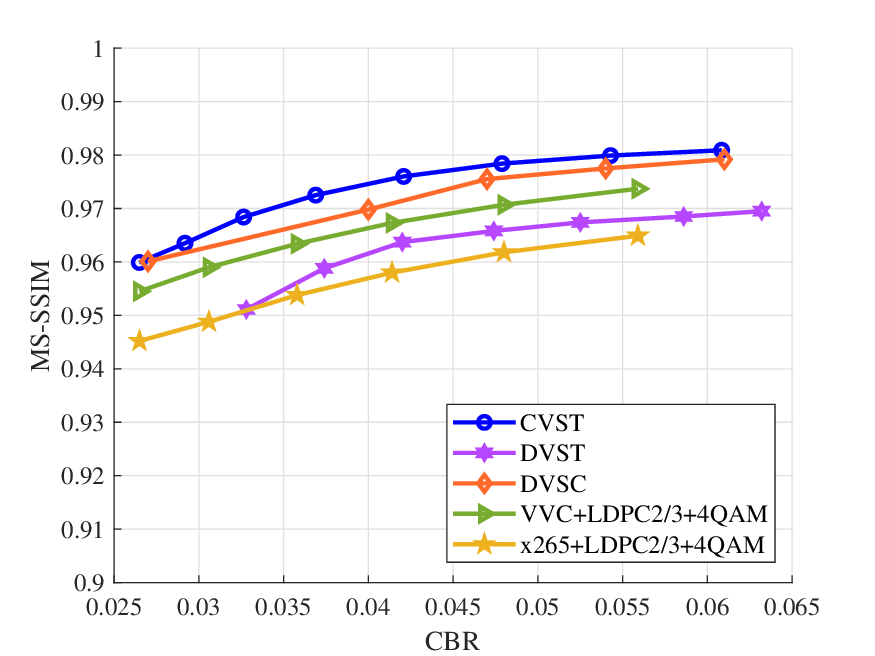}
	\caption{The MS-SSIM performance. (HEVC Class C)}
	\label{fig_8}
\end{figure}
\subsubsection{Performance for semantic-relevant evaluation indexes}
\begin{figure}[htbp]
	\centering
	\includegraphics[width=3.6in]{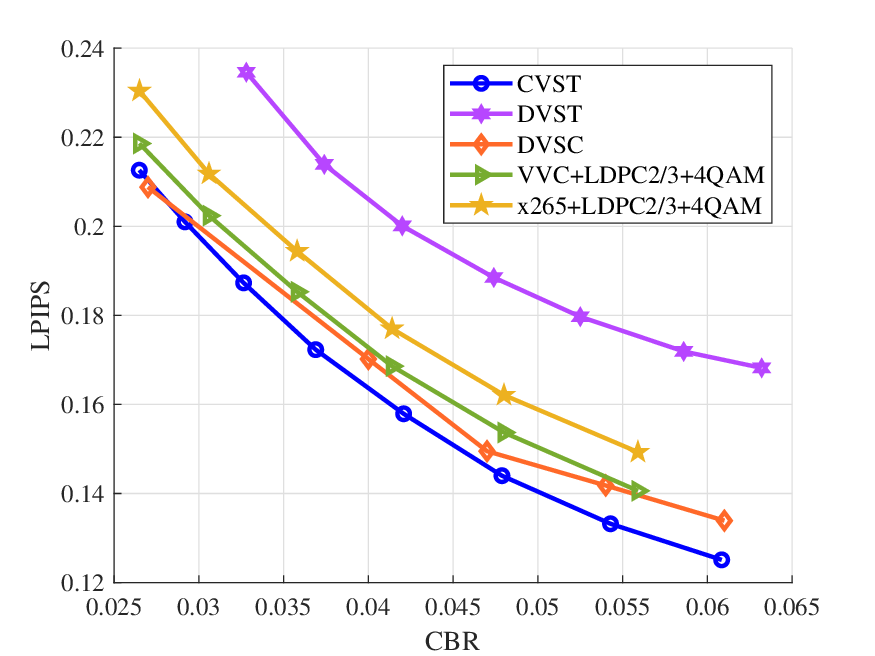}
	\caption{The LPIPS performance. (HEVC Class C)}
	\label{fig_9}
\end{figure}

To further evaluate the human-perceptual quality of the reconstructed videos, we employ advanced perception-oriented metrics, including MS-SSIM and LPIPS. As illustrated in Fig. \ref{fig_8} and Fig. \ref{fig_9}, CVST consistently surpasses both DVST and DVSC, demonstrating its superior ability to retain perceptual quality. This advantage stems from CVST's learned context, which effectively captures and preserves the essential semantic information embedded within the original videos, thereby maintaining finer visual details and structural integrity even under challenging wireless transmission conditions. In contrast, the traditional SSCC scheme remains susceptible to the cliff effect, where video quality degrades precipitously once channel conditions fall below a critical threshold. These collective results affirm that the channel and rate-aware CVST enables robust and human-friendly reconstructed videos tailored for practical wireless environments.
\begin{figure}[htbp]
	\centering
	\includegraphics[width=3.6in]{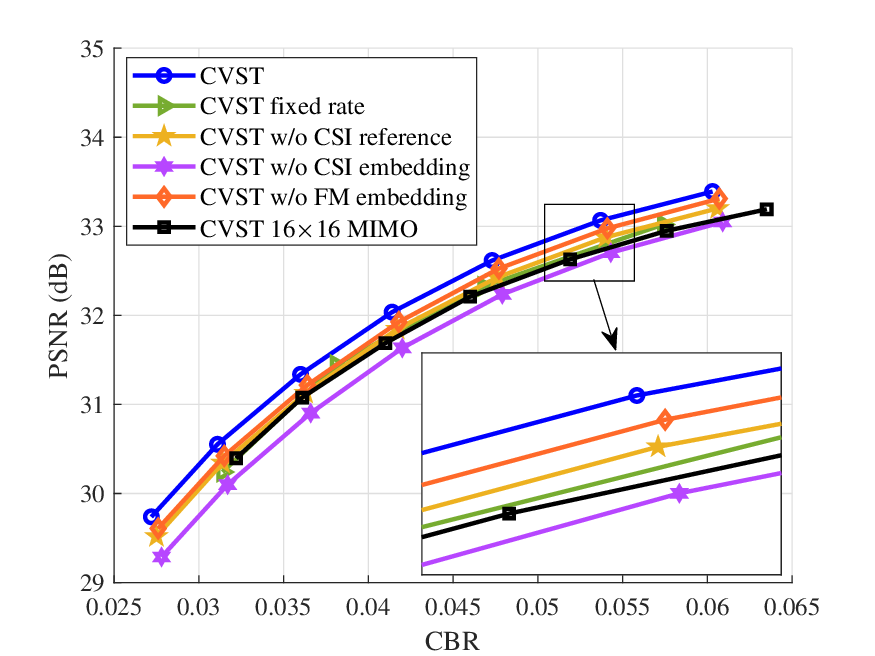}
	\caption{Ablation studies for different designed modules.}
	\label{fig_10}
\end{figure}

\begin{table}[htbp]
	\centering
	\caption{Evaluation of different $\eta$ ranges. (CBR=0.036, SNR=9 dB)}
	\label{table3}
	
	\begin{tabular}{|c|c|c|c|c|}  
		\hline 
		& & & &\\[-6pt] 
		&$[\eta^c_{\mathrm{min}},\eta^c_{\mathrm{max}}]$&$[\eta^v_{\mathrm{min}},\eta^v_{\mathrm{max}}]$&PSNR (dB)&MS-SSIM \\
		\hline
		& & & &\\[-6pt]  
		1&$[0.25,0.55]$&$[0.07,0.15]$&31.43&0.9716 \\
		\hline
		& & & &\\[-6pt]  
		2&$[0.4,0.8]$&$[0.07,0.15]$&31.22&0.9701 \\
		\hline
		& & & &\\[-6pt]  
		3&$[0.2,0.5]$&$[0.2,0.5]$&31.09&0.9690 \\
		\hline
		& & & &\\[-6pt] 
		4&$[0.1,0.4]$&$[0.2,0.5]$&30.98&0.9681 \\
		\hline
	\end{tabular}
\end{table}

\subsubsection{Ablation Studies for the Proposed Modules}
To validate the contribution of each designed component, ablation studies for various designed modules are provided in Fig. \ref{fig_10}, where `CVST w/o CSI reference' denotes the absence of CSI references in the entropy coding process, `CVST w/o CSI embedding' and `CVST w/o FM embedding' denote the exclusion of the context-channel correlation map and the feature modulation terms, respectively, as embedded SEI in the MR-VLRC module; `CVST fixed rate' indicates the removal of variable rate coding by fixing the feature modulation term, requiring separate models for each rate point; `CVST 16$\times$16 MIMO' refers to the 16$\times$16 MIMO antenna type. The results demonstrate that both CSI and FM embeddings pose positive effects to entropy coding by providing essential SEI for CSI-aware and rate-adaptive transmission. Moreover, the variable rate approach in CVST performs comparably to training multiple fixed-rate models, confirming its efficacy in enabling flexible and adaptive transmission in MIMO wireless scenarios. Finally, the complexity scaling with MIMO dimensions does not bring much performance degradation, which verify the potential feasibility for large scale MIMO channels.

As shown in Tab. III, $\eta$ ranges are varied to evaluate the effect of both motion vectors and context brought to wireless video transmission. The first range pair is our CVST's setting for all other experiments, which takes the importance of context and effect of motion vectors into consideration. While for other configurations, the second range pair put too much attention to the context information, neglecting the motion changes with motion-heavy clips. The second pair put motion vectors with equal importance to the context, which degrades the performance due to the overlook of rich context, let alone the fourth pair with inverse rate allocation weight. The ablation experiment shows the balance between context and motion vectors in terms of rate allocation.

\subsubsection{Evaluation for the imperfect CSI}
\begin{figure}[htbp]
	\centering
	\includegraphics[width=3.4in]{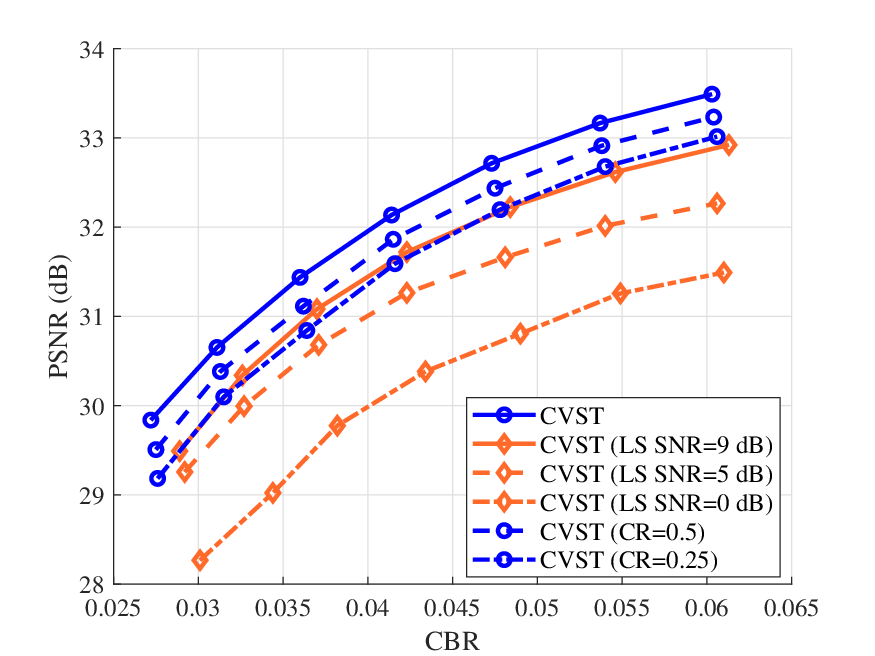}
	\caption{Evaluations for the imperfect CSI. (SNR = 9 dB in default)}
	\label{fig_11}
\end{figure}

In previous results, we evaluate the performance of CVST and all other benchmarks based on perfect CSI, which is an ideal condition. As such, more practical condition with imperfect CSI is conducted. Least square (LS) channel estimation is adapted to evaluate the estimation loss while DNN-based network is employed for CSI feedback with different feedback CSI compression ratio (CR). As shown in Fig. \ref{fig_11}, imperfect CSI leads to a noticeable performance loss. Since LS is a relatively simple estimator, estimation errors degrade the end-to-end transmission quality of CVST. Moreover, as the SNR decreases, the performance gap between LS-based CSI and perfect CSI becomes larger; for example, we observe an approximately 2 dB PSNR loss for CVST with LS at SNR = 0 dB compared with the perfect-CSI case at SNR = 9 dB. Then we introduce the CSI feedback into CVST, the CR=0.25, 0.5 show some performance loss for CVST due to the imperfect feedback CSI, about 0.7 dB for the CR=0.25. In conclusion, for the imperfect CSI, channel estimation error seems to dominate the degradation. However, with the deployment of advanced channel estimation and CSI feedback techniques, nearly perfect performance can be achieved with imperfect CSI.

\begin{figure}[htbp]
	\centering
	\includegraphics[width=3.4in]{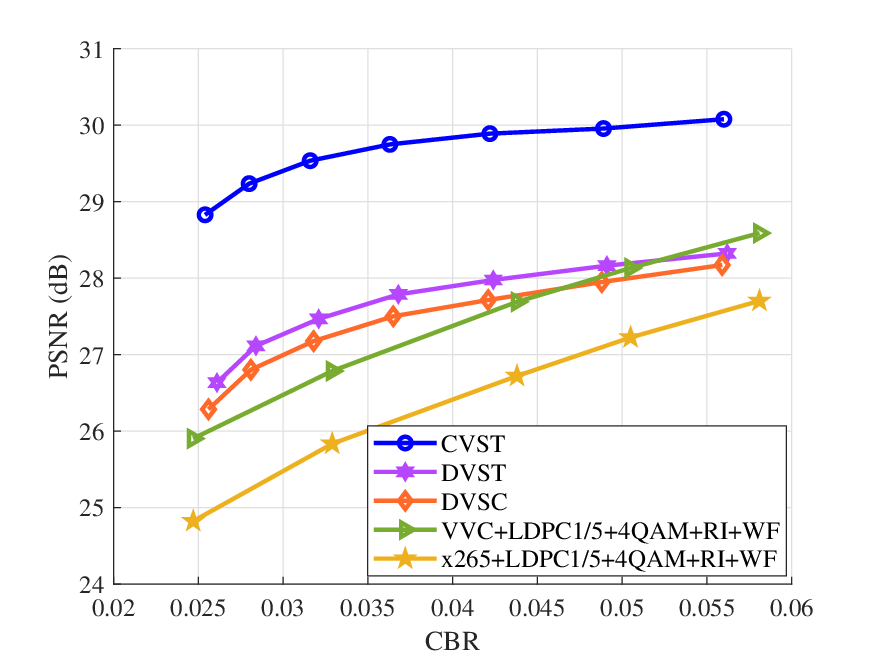}
	\caption{Performance under 3GPP CDL Channels.}
	\label{fig_12}
\end{figure}

\begin{figure*}[htbp]
	\centering
	\includegraphics[width=6.0in]{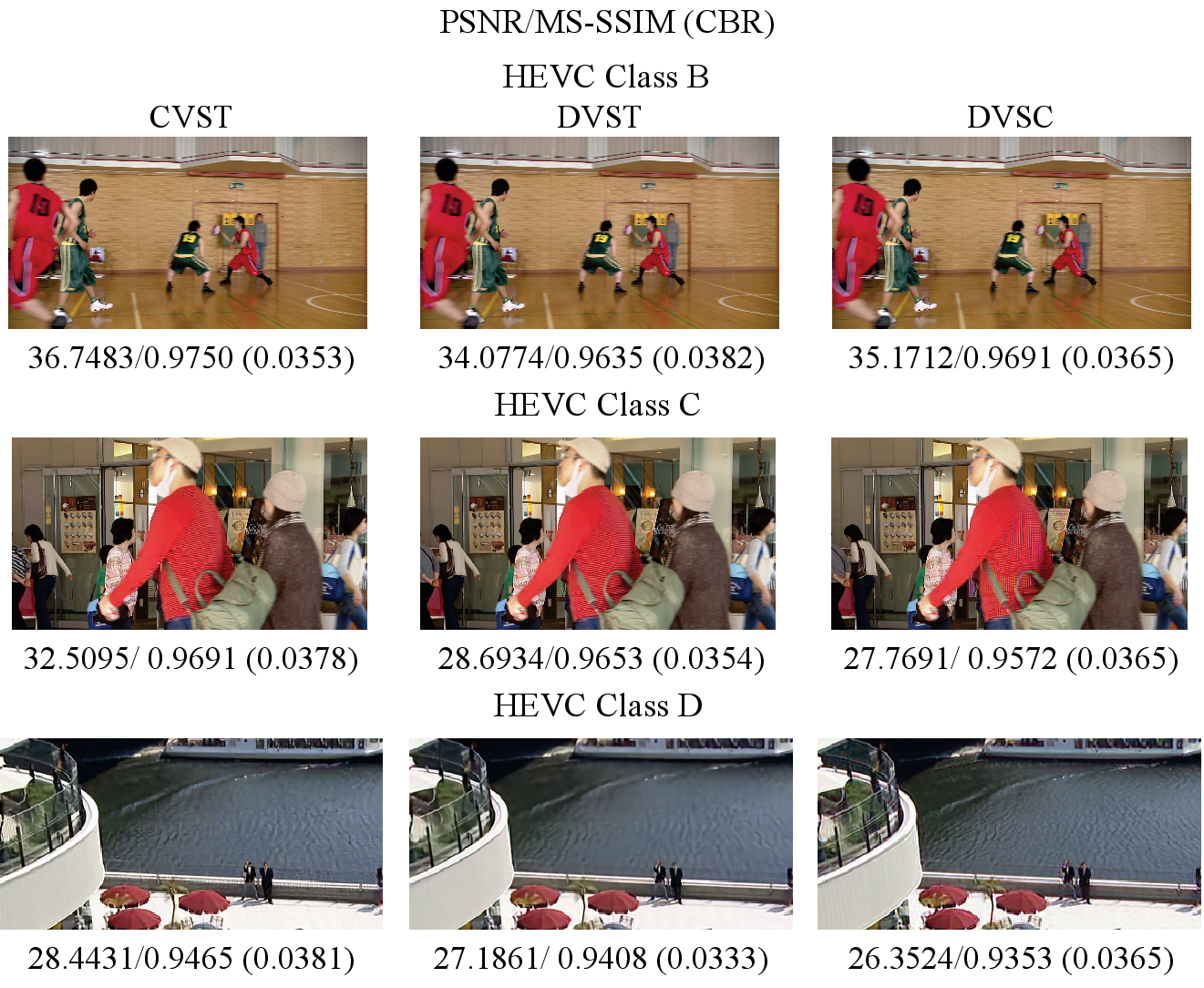}
	\caption{Visualized results for the CVST and other benchmarks for HEVC Class B, Class C, and Class D. (SNR = 8 dB)}
	\label{fig_13}
\end{figure*}

\subsubsection{Performance under 3GPP CDL Channels}

\begin{table}[htbp]
	\centering
	\caption{MIMO CDL channel-related Configuration}
	\label{table4}
	
	\begin{tabular}{|c|c|c|c|}  
		\hline
		& & &\\[-6pt]  
		MIMO&8$\times$8&Subcarrier&64 \\
		\hline
		& & &\\[-6pt]  
		Channel Model&\thead{3GPP 38.901 \\ CDL-A}&Speed&7.2 km/h \\
		\hline
		& & &\\[-6pt] 
		Carrier Frequency&2.6GHz&Direction&uplink\\
		\hline
	\end{tabular}
\end{table}

Then, we further evaluate the performance of CVST under 3GPP MIMO clustered delay line (CDL) channels at SNR = 16 dB. Since CDL channels are time-correlated and frequency-selective due to multipath propagation, we adopt an OFDM-based transmission model to capture the resulting frequency selectivity. The corresponding MIMO-OFDM settings are summarized in Tab. IV, where each subcarrier experiences an 8$\times$8 MIMO channel matrix. For simplicity and to limit signaling complexity, we use the average MIMO channel response across all subcarriers as the CSI input for constructing the context–channel correlation map $\mathbf{m}_t$. As shown in Fig. \ref{fig_13}, CVST still achieves satisfactory performance under multi-path CDL channels compared to DVSC and DVST. For the traditional SSCC schemes, CVST also achieves obvious performance gain due to the context video transmission structure and VLRC adaptive designs. In conclusion, CVST is competent for wireless video transmission under time-correlated multi-path channels.

\subsubsection{Visualization Results for the Wireless Video Transmission}
Fig. \ref{fig_13} visualizes reconstructed video results from the HEVC Class B, C, and D datasets with diverse contents and resolutions, comparing CVST against related DL-based benchmarks. Among other benchmarks, CVST delivers superior visual fidelity, particularly outperforming DVST and DVSC. The reconstructed frames confirm CVST’s ability to maintain robust visual quality under wireless transmission constraints.

\subsubsection{Complexity Analysis}
\begin{table}[htbp]
	\centering
	\caption{Evaluation of complexity and computation cost.}
	\label{table5}
	
	\begin{tabular}{|c|c|c|c|}  
		\hline 
		& & &\\[-6pt] 
		Metric&FLOPs (G)&Throughput&Parameters (M) \\
		\hline
		& & &\\[-6pt]  
		CVST&411.37&1.17&24.96 \\
		\hline
		& & &\\[-6pt]  
		DVST&556.05&1.28&5.81 \\
		\hline
		& & &\\[-6pt] 
		DVSC &372.54&1.74&15.74\\
		\hline
	\end{tabular}
\end{table}

Finally, to evaluate the practical deployability of CVST, we analyze its computational cost relative to DVSC and DVST. FLOPs quantify computational complexity, throughput measures inference speed (average frames per second on HEVC Class C), and the number of parameters reflects model size. The inference batch size is set to 1 to emulate per-frame reconstruction. As shown in Tab. \ref{table5}, CVST achieves competitive throughput while delivering substantially better reconstruction quality. This efficiency is enabled by the lightweight CNN-based context-coding backbone and the checkerboard entropy-coding structure, which reduces sequential dependencies and enables fine-grained parallelism during entropy modeling. Although the proposed multi-reference entropy coding introduces additional parameters, CVST retains a practical inference speed. In contrast, DVSC attains higher throughput largely because it does not include entropy coding, which comes at the cost of reduced performance; DVST incurs higher FLOPs due to its Transformer-based backbone and the autoregressive entropy model. Overall, these results confirm that CVST achieves a favorable performance–complexity tradeoff and is amenable to practical deployment, with further acceleration possible via parallel implementation and GPU optimization.

\section{Conclusion}
In this paper, we proposed CVST, which integrates context-based transmission with multi-reference entropy coding, generating a context-channel correlation map that is adaptively embedded as SEI. This integration enables transmission performance that remains aware of MIMO CSI. A key innovation of our framework is a checkerboard-based feature modulation method, which allows a single trained CVST model to support a wide range of CBRs. Extensive experiments demonstrate that CVST achieves highly effective and robust performance for variable length and variable rate video transmission across diverse channel conditions. In the future, we will explore various channel types and conditions along with the multi-user scenarios to further improve CVST into practical use.

\fontsize{8pt}{10pt}\selectfont

\end{document}